\documentclass[aps, pra, showpacs, twocolumn, 10pt, superscriptaddress, nofootinbib]{revtex4-1}

\makeatletter
\def\@bibdataout@aps{%
 \immediate\write\@bibdataout{%
  @CONTROL{%
   apsrev41Control,author="08",editor="1",pages="0",title="0",year="1"%
  }%
 }%
 \if@filesw
  \immediate\write\@auxout{\string\citation{apsrev41Control}}%
 \fi
}%
\makeatother 

\usepackage{bbm, amsmath, amssymb, amsthm, bm,textcomp, nicefrac,geometry}
\usepackage{graphicx,epstopdf,color,verbatim,enumitem}
\usepackage{commath}
\usepackage{tabu}
\usepackage{times}
\usepackage[dvipsnames]{xcolor}
\usepackage{dsfont}
\usepackage[none]{hyphenat}
\setcitestyle{numbers,square,citesep={,\kern-.24em}}
\usepackage{hyperref}
\hypersetup{
    colorlinks=true,
    linkcolor=blue,
    filecolor=magenta,
    urlcolor=blue,
    citecolor=blue
}

\DeclareMathOperator{\tr}{{tr}}

\newcounter{observ}
\newenvironment{observ}{\refstepcounter{observ}\textit{Observation \theobserv}.}

\newenvironment{proo}{\textit{Proof.}}{\hfill$\blacksquare$}

\newcommand{\ket}[1]{\ensuremath{\vert #1 \rangle}}

\newcommand{\braket}[2]{\ensuremath{\langle #1 \vert #2 \rangle}}
\newcommand{\braOket}[3]{\ensuremath{\langle #1 \vert #2 \vert #3 \rangle}}
\newcommand{\ketbra}[2]{\ensuremath{\vert #1 \rangle \! \langle #2 \vert}}
\newcommand{\meanO}[1]{\ensuremath{\langle #1 \rangle}}
\newcommand{\ver}[2]{\ensuremath{\genfrac{}{}{0pt}{}{#1}{#2}}}
\newcommand{\trcua}[1]{\ensuremath{\tr \lcua #1\rcua}}

\newcommand{\bs}[1]{\ensuremath{\boldsymbol{#1}}}
\newcommand{\ex}[1]{\ensuremath{\langle{#1}\rangle}}
\newcommand{\varian}[1]{\ensuremath{(\Delta #1 )^{ 2}}}
\newcommand{\varinv}[1]{\ensuremath{(\Delta #1 )^{-2}}}
\newcommand{\qfif}[1]{\ensuremath{\qfi[{\textstyle #1}]}}
\newcommand{\dicke}[1]{\ensuremath{{\text{D}_{#1}}}}

\newcommand{\units}[2]{\ensuremath{{#1}{\text{~#2}}}}

\def\be{\begin{equation}}
\def\ee{\end{equation}}
\def\bea{\begin{eqnarray}} 
\def\eea{\end{eqnarray}}
\def\bse{\begin{subequations}}
\def\ese{\end{subequations}}
\def\mtxid{\mathbbm{1}}
\def\qfi{\mathcal{F}_{\text{Q}}}
\def\ghz{{\text{GHZ}}}
\def\lpar{\left(}
\def\rpar{\right)}
\def\lcua{\left[}
\def\rcua{\right]}
\def\lcor{\left\{}

\begin{document}
\setlength{\abovedisplayskip}{0.5em}
\setlength{\belowdisplayskip}{0.5em}

\title{Precision bounds for gradient magnetometry with atomic ensembles}
\author{Iagoba Apellaniz}\email{iagoba.apellaniz@gmail.com}
\affiliation{Department of Theoretical Physics,
University of the Basque Country UPV/EHU, P. O. Box 644, E-48080 Bilbao, Spain}
\author{I\~nigo Urizar-Lanz}
\affiliation{Department of Theoretical Physics,
University of the Basque Country UPV/EHU, P. O. Box 644, E-48080 Bilbao, Spain}
\author{Zolt\'an Zimbor\'as}
\affiliation{Department of Theoretical Physics,
University of the Basque Country UPV/EHU, P. O. Box 644, E-48080 Bilbao, Spain}
\affiliation{Dahlem Center for Complex Quantum Systems, Freie Universit\"at Berlin, 14195 Berlin, Germany}
\affiliation{Wigner Research Centre for Physics, Hungarian
Academy of Sciences, P.O. Box 49, H-1525 Budapest, Hungary}
\author{Philipp Hyllus}
\affiliation{Department of Theoretical Physics,
University of the Basque Country UPV/EHU, P. O. Box 644, E-48080 Bilbao, Spain}
\author{G\'eza T\'oth}
\email{toth@alumni.nd.edu}
\homepage{http://www.gtoth.eu}
\affiliation{Department of Theoretical Physics,
University of the Basque Country UPV/EHU, P. O. Box 644, E-48080 Bilbao, Spain}
\affiliation{Wigner Research Centre for Physics, Hungarian
Academy of Sciences, P.O. Box 49, H-1525 Budapest, Hungary}
\affiliation{IKERBASQUE, Basque Foundation for Science, E-48013 Bilbao, Spain \vspace{5pt}}

\makeatletter
  \def\Dated@name{}
\makeatother
\date{Received 7 July 2017; published 8 May 2018}

\begin{abstract}

We study gradient magnetometry with an ensemble of atoms with arbitrary spin.
We calculate precision bounds for estimating the gradient of the magnetic field based on the quantum Fisher information.
For quantum states that are invariant under homogeneous magnetic fields, we need to measure a single observable to estimate the gradient.
On the other hand, for states that are sensitive to homogeneous fields, a simultaneous measurement is needed, as the homogeneous field must also be estimated.
We prove that for the cases studied in this paper, such a measurement is feasible.
We present a method to calculate precision bounds for gradient estimation with a chain of atoms or with two spatially separated atomic ensembles.
We also consider a single atomic ensemble with an arbitrary density profile, where the atoms cannot be addressed individually, and which is a very relevant case for experiments.
Our model can take into account even correlations between particle positions.
While in most of the discussion we consider an ensemble of localized particles that are classical with respect to their spatial degree of freedom,  we also discuss the case of gradient metrology with a single Bose-Einstein condensate.

\vspace{1em}\noindent
DOI: \href{https://doi.org/10.1103/PhysRevA.97.053603}{10.1103/PhysRevA.97.053603}
\end{abstract}

\vspace*{1.5em}

\maketitle

\section{Introduction}
Metrology plays an important role in many areas
of physics and engineering \cite{Glaser2010}.
With the development of experimental techniques,
it is now possible to realize metrological tasks
in physical systems that cannot
be described well by classical physics and instead quantum mechanics must be
used for their modeling.
Quantum metrology \cite{Giovannetti2004, Giovannetti2006, Paris2009, Gross2012}
is the novel field that is concerned with
metrology using such quantum mechanical systems.

One of the basic tasks of quantum metrology is magnetometry with an ensemble
of spin-$j$ particles.
Magnetometry with a completely
polarized state works as follows.
The total spin of the ensemble is
rotated by a homogeneous magnetic field perpendicular to it.
We would like to estimate the rotation angle or phase $\theta$ based on some
measurement; this phase parameter can then be used to obtain the field strength.
To determine the rotation angle, one needs, for instance, to measure a spin component perpendicular to the mean spin.

Up to now, it looks as if the total spin behaves like a clock arm and
its position tells us the value of $\theta$ exactly.
At this point one has to remember that we have an ensemble of $N$
particles governed by quantum mechanics,
and the uncertainty of the spin component
perpendicular to the mean spin can never be zero.
Hence, simple calculation shows that the scaling of the precision of the phase estimation is $\varinv{\theta}\,{\sim}\,N$, which is called \emph{shot-noise scaling} \cite{Giovannetti2004, Giovannetti2006, Paris2009, Gross2012}.
However, spin squeezing \cite{Kitagawa1993,
Wineland1994, Sorensen2001, Ma2011, Kuzmich1998} can decrease the uncertainty of
one of the components  perpendicular to the mean
spin and this can be used to increase the
precision of the measurements \cite{Kuzmich1998}.
While it is possible to surpass the shot-noise limit, for the case of a linear Hamiltonian \cite{Giovannetti2004, Giovannetti2006, Paris2009, Gross2012},
no quantum state can have a better scaling in the precision than
$\varinv{\theta}\,{\sim}\,N^{2}$,
called \emph{Heisenberg scaling}.

In recent years, quantum metrology has been applied in many scenarios, from
atomic clocks \cite{Louchet-Chauvet2010, Borregaard2013, Kessler2014a} and
precision magnetometry \cite{Wasilewski2010,
Eckert2006, Wildermuth2006, Wolfgramm2010, Koschorreck2011, Vengalattore2007, Zhou2010} to gravitational wave
detectors \cite{Schnabel2010,TheLIGOScientificCollaboration2011, Demkowicz-Dobrzanski2013}. So far, most of the attention has
been paid to the problem of estimating a single parameter.
The case of multiparameter estimation for
quantum systems is much less studied, possibly, since
it can be more complicated due to the noncommutative nature of the problem \cite{kolodynski2010,Crowley2014,Monras2011, Vaneph2013, Knysh2013, Matsumoto2002, Baumgratz2016, Szczykulska2016, Marzolino2013, Marzolino2015Erratum, Humphreys2013, Skotiniotis2015, Knott2016,Pezze2017Optimal,Ciampini2016Quantum}.

\begin{figure}[h]
  \begin{center}
    \includegraphics[width=200pt]{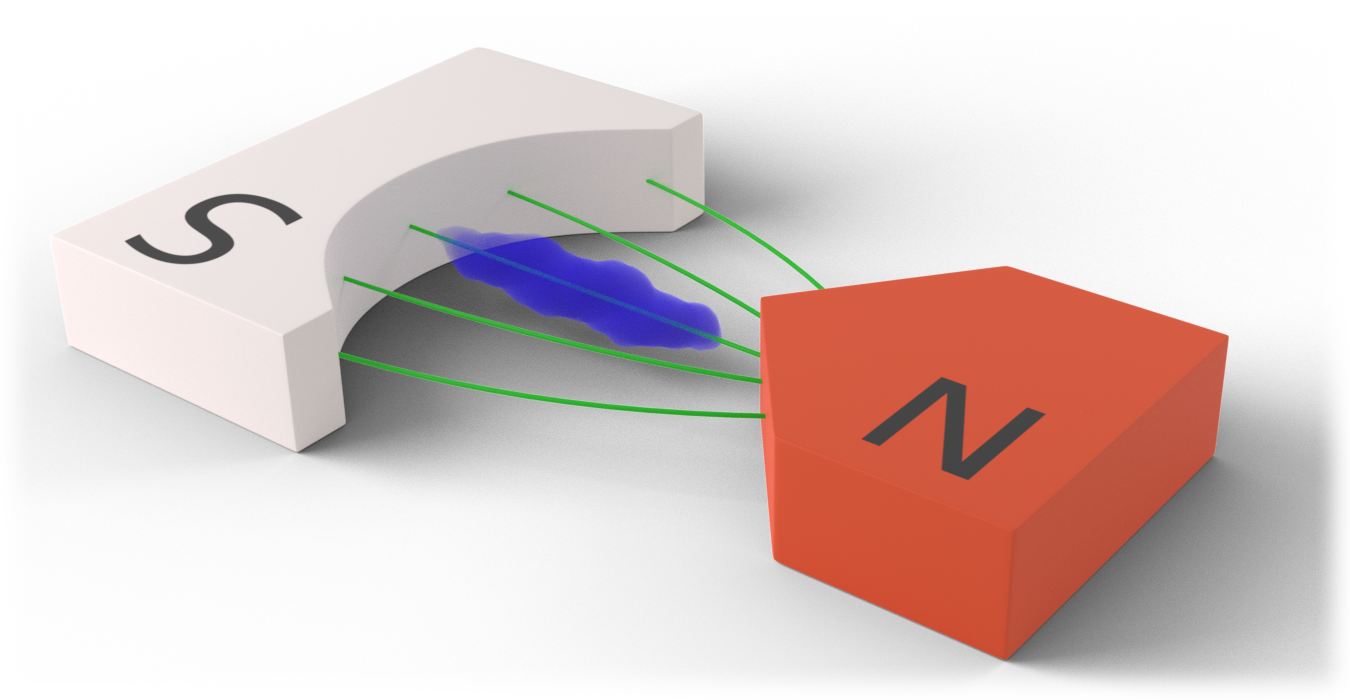}
    \caption{Schematic representation of an atomic ensemble (blue cloud) placed in a magnetic field (green lines) in a Stern-Gerlach apparatus. From the final state the gradient of the field can be estimated. Note that the field intensity changes along the cloud of atoms, while the direction is always the same from north (N) to south (S) within the cloud. For an easier presentation, a setup is shown where the direction of the magnetic field is parallel to the cloud, however, this does not have always to be the case. }\label{fig:cloud-in-sg}
  \end{center}
\end{figure}

In this paper, we compute precision bounds for the estimation of the magnetic field gradient (see Fig.~\ref{fig:cloud-in-sg}).
In general, in order to achieve these bounds, an estimate of the constant (homogeneous) part of the field is required.
Hence, we have to use the formalism of multiparameter estimation.
Magnetometry of this type  can be realized with differential interferometry with two particle ensembles, which has raised a lot of attention in quantum metrology \cite{Landini2014, Eckert2006, Stockton2007, Durfee2006, Snadden1998, Fixler2007, Altenburg2016}.
Another possibility is considering spin chains, which can be relevant in trapped cold ions or optical lattices of cold atoms, where we have individual access to the particles \cite{Urizar-Lanz2013, Zhang2014, Ng2014}.

Finally,
gradient magnetometry can be carried out using a single atomic cloud,
which is very relevant from the point of view of cold gas experiments.
One can consider both atomic clouds of localized particles, as
well as Bose-Einstein condensates.
While most works in magnetometry with a single
ensemble focus only on the determination of the strength and direction of
the magnetic field,
certain measurement schemes for the gradient have already
been proposed and tested experimentally. Some schemes
use an imaging of the
ensemble with a high spatial resolution.
\pagebreak
They do not count as single-ensemble methods in the sense
we use this expression in our paper since in this case not only
collective observables are measured  \cite{Vengalattore2007, Zhou2010,Koschorreck2011}.
There is a method based on collective measurements of the
spin length of a fully polarized ensemble given in Ref.~\cite{Behbood2013}.
There is also a scheme based on many-body singlet states described
in Ref.~\cite{Urizar-Lanz2013}.

We use the quantum Fisher information (QFI) and the Cram\'er-Rao (CR) bound in our derivations \cite{Paris2009,Braunstein1994,Holevo1982,Helstrom1976,Petz2002,Petz2008}.
Due to this, our calculations are generally valid for any measurement, thus they are
relevant to many recent experiments \cite{Wasilewski2010,
Eckert2006,Wildermuth2006, Wolfgramm2010,Koschorreck2011,Vengalattore2007,Zhou2010,Behbood2013}.
We note that in the case of the spin singlet, our precision bounds are saturated
by the metrological scheme presented in Ref.~\cite{Urizar-Lanz2013}.

We can also connect our results to entanglement theory \cite{Werner1989,Horodecki2009,Guehne2009}.
We find that the shot-noise scaling cannot be surpassed with separable
states, while the Heisenberg scaling can be reached with entangled states.
However, 
the shot-noise scaling can be surpassed only if
the particle positions are correlated, which is the case, for instance,
if the particles attract each other.

Next, we present the main characteristics of our setup.
For simplicity, as well as following  recent experiments (e.g., Ref.~\cite{Koschorreck2011}),
we consider an ensemble of spin-$j$ particles
placed in a one-dimensional arrangement.
The atoms are then situated along the $x$ axis with $y=z=0.$
We assume that we have particles
that behave classically with respect to their spatial state.
That is, they cannot be in a superposition of being at two different places.
On the other hand, they have internal degrees of freedom, their spin, which is quantum.
This is a very good description to many of the cold gas experiments.

Based on these considerations, we assume that the state is factorizable into
a spatial part and a spin part as
\be
    \label{eq:state-internal-external}
    \varrho=\varrho^{(\text{x})}\otimes\varrho^{(\text{s})},
\ee
where the internal state is decomposed in its eigenbasis as
\be\label{eq:rhoeigendecomp}
\varrho^{(\text{s})} =\sum_{\lambda} p_\lambda \ketbra{\lambda}{\lambda}.
\ee
For the spatial part defined in the continuous Hilbert space, we assume that it can be modeled by an incoherent mixture of pointlike particles as
\be
    \label{eq:thermal-state}
    \varrho^{(\text{x})}=\int \frac{P(\bs{x})}{\braket{\bs{x}}{\bs{x}}} \ketbra{\bs{x}}{\bs{x}} \dif{\bs{x}},
\ee
where $\bs{x} =(x_1,x_2,\dots,x_N)$ is a vector which collects all the particle positions, $P(\bs{x})$ is the spatial probability distribution function of the atoms, and $\dif{\bs{x}}$ denotes $\dif{x_1} \dif{x_2} \cdots \dif{x_N}.$
Note that the spatial part \eqref{eq:thermal-state} is diagonal in the position eigenbasis, which simplifies considerably our calculations (see Appendix~\ref{app:thermal-state} for more details).
During the evolution of the state, correlations might arise between the internal and the spatial parts and the product form \eqref{eq:state-internal-external} might not be valid to describe the evolution of the system.

First, we consider spin chains and two particle ensembles at different places.
The gradient measurement with two ensembles is essentially based on the idea that the gradient is just the difference between two measurements at different locations. With these systems, it is possible to reach the Heisenberg scaling.

We also examine in detail the case of a single atomic en- semble.
Since in such systems the atoms cannot be individually
addressed, we assume that the quantum state is permutationally invariant (PI).
We show that for states insensitive to the homogeneous magnetic field,
one can reduce the problem to a one-parameter estimation scenario.
Single-ensemble measure- ments have certain advantages because
the spatial resolution can be higher and the experimental requirements are
smaller since only a single ensemble must be prepared.

For completeness, we mention the case of Bose-Einstein condensates (BEC). The spatial state in this case is pure
\be
\label{eq:BEC}
\varrho^{(\text{x})}_{\rm BEC}=(\ketbra{\Psi}{\Psi})^{\otimes N},
\ee
where $\vert\Psi\rangle$ is the spatial state of a single particle.
Hence, the spatial state is delocalized and it is not an incoherent mixture of various eigenstates of $x$. While we do not consider such systems in detail, our formalism could be used to model them.

We now outline the model we use to describe the interaction of the particles with the magnetic field.
The field at the atoms is given as
\begin{equation}
  \bs{B}(x,0,0)=\bs{B}_0 +x \bs{B}_1 + \mathcal{O}(x^2),
\end{equation}
where we neglect the terms of order two or higher, and where $\mathcal{O}(\xi)$ is the usual Landau notation to describe the asymptotic behavior of a quantity, in this case for small $\xi$.
We consider the magnetic field pointing in the $z$ direction, hence,
$\bs{B}_0=B_0 (0,0,1)$ and
$\bs{B}_1=B_1 (0,0,1)$.
For this configuration, due to the Maxwell equations, with no currents or changing electric fields, we have
\be
  \text{div} \, \bs{B} = 0,  \quad
  \text{curl}\, \bs{B} = (0,0,0).
\ee
This implies $\sum_{l=x,y,z} \partial B_l /\partial l=0$ and $ \partial B_l /\partial m- \partial B_m /\partial l=0$ for $l\ne m.$
Thus, the spatial derivatives of the field components are not independent of each other.
In this paper, however, we consider an elongated trap.
In the case of such a quasi-one- dimensional atomic ensemble,
only the derivative along the axis of the trap has an influence on the quantum
dynamics of the atoms or a double-well experiment.

We determine the precision bounds for the estimation of the magnetic field gradient $B_1$.
We calculate how the precision scales with the number of particles.
We compare systems\break
with an increasing particle number, but of the same size.
As discussed later, if we follow a different route, we can obtain results
that can incorrectly be interpreted as reaching the Heisenberg limit, or even a
super-Heisenberg scaling.

The angular momentum of an individual atom is coupled to the magnetic field, yielding the following interaction term:
\be\label{eq:int}
    h^{(n)}=\gamma B_z^{(n)} \otimes j_z^{(n)},
\ee
where the operator $B_z^{(n)}=B_0+B_1\hat{x}^{(n)}$ acts on
the spatial part of the Hilbert space and $\hat{x}^{(n)}$ is the position operator of a single particle. Moreover, $j_z^{(n)}$ is a single-particle spin operator, acting on the spin part of the Hilbert space.
Finally, $\gamma = g \mu_{\text{B}}$ where $g$ is the gyromagnetic factor and $\mu_{\text{B}}$ corresponds to the Born magneton, and
we set $\hbar=1$ for simplicity.
We use the ``$\;\hat{\,}\;$'' notation to distinguish the operator $\hat{x}$ from the coordinate $x.$
Later, we will omit it for simplicity.
The Hamiltonian of the entire system is just the sum  of all two-particle interactions of the type Eq.~\eqref{eq:int} and can be written as
\be
\label{eq:Htot}
H = \gamma \sum_{n=1}^N B_z^{(n)} \otimes j_z^{(n)}.
\ee
Equation~\eqref{eq:Htot} generates the time evolution  of the atomic ensemble.

One could include also the kinetic energy in the Hamiltonian. Such an extra term causes that the gradient field pushes atoms in state $\ket{0}$ into one direction, while atoms in state $\ket{1}$ into the other direction. In our work, we do not take into account this effect. Moreover, we do not include in the model the initial thermal dynamics of the particles. Both of these effects are negligible in a usual setup, as shown in Appendix~\ref{app:experimental-suport-for-omitting-bireANDthermal}.

We calculate lower bounds on the precision of estimating $B_1$ based on a measurement on the state after it passed through the unitary dynamics $U=\exp(-iHt)$, where $t$ is the time spent by the system under the influence of the magnetic field.
The unitary operator can be rewritten as
\be
    \label{eq:unitary-operator}
    U=e^{-i \lpar b_0 H_0 + b_1 H_1 \rpar},
\ee
where the $b_i=\gamma B_i t$.
The generator describing the effect of the homogeneous field is  given as
\be
    \label{eq:homogeneous-generator}
    H_0=\sum_{n=1}^N j_z^{(n)} = J_z,
\ee
while the generator describing the effect of the gradient is
\be
    \label{eq:gradient shift generator}
    H_1=\sum_{n=1}^N x^{(n)}j_z^{(n)}.
\ee
We omit $\otimes$ and the superscripts (x) and (s) for simplicity, and use them only if it is necessary to avoid confusions.

The operators $H_{0}$ and $H_{1}$ commute with each other.
However, it is not necessarily
true that the operators we have to measure to estimate $b_0$ or $b_1$
can be simultaneously measured.
The reason for that is that both operators to be measured
act on the same atomic ensemble.
If the measurement operators do not commute with each other, then the precision bound obtained from the theory of QFI cannot necessarily be reached.
For the particular cases studied in this paper, we prove that a simultaneous measurement to estimate both the homogeneous and the gradient parameter can be carried out (see Appendix~\ref{app:compatibility-of-measurements}).
On the other hand, in schemes in which the gradient is calculated
based on measurements on two separate atomic ensembles or
different atoms in a chain, the measuring operators can always commute
with each other \cite{Wasilewski2010, Eckert2006, Zhang2014}.

The paper is organized as follows. In Sec.~\ref{sec:cramer-rao bounds}, general precision bounds for the estimation of the gradient of the magnetic field are presented.
In Sec.~\ref{sec:twin cloud systems}, we compute precision bounds for relevant spatial configurations appearing in cold atom physics such as spin chains and  two ensembles spatially separated from each other.
In Sec.~\ref{sec:single cloud systems}, we consider a single atomic ensemble in a PI state and we
calculate the precision bounds for various quantum states,
such as the singlet spin state
or the totally polarized state.
In Sec.~\ref{sec:bec}, we consider Bose-Einstein condensates.

\section{Precision bounds\\ for estimating the gradient}
\label{sec:cramer-rao bounds}

In this section, we show how the QFI helps us to obtain the bound on the precision of the gradient estimation.
First, we discuss gradient magnetometry using quantum states that are insensitive to homogeneous fields. In this case, we need to estimate only the gradient and do not have to know the homogeneous field. Hence, this case corresponds to a single-parameter estimation problem.

Then, we discuss the case of quantum states sensitive to homogeneous fields.
Even in this case, we are interested only in the gradient, and we do not aim at estimating the homogeneous field.
In spite of this, gradient estimation with such states is a two-parameter estimation task. We introduce the basics of multi-parameter quantum metrology, and we adapt that formalism to our problem.
We also show that the precision bound obtained does not change under spatial translation, which will be used later to simplify our calculations.
In Appendix~\ref{app:compatibility-of-measurements}, we show that even the precision bounds for states sensitive to the homogeneous field, appearing in this paper, are saturable.

Next, we summarize important properties of the QFI used throughout this paper (for reviews,
see Refs.~\cite{Giovannetti2004,Giovannetti2011,Paris2009,Demkowicz-Dobrzanski2014Quantum,Pezze2014Quantum,Toth2014,Pezze2016Non-classical,
2017BraunQuantum_arxiv,
Petz2008}).
Let us consider a quantum state with the eigendecomposition
\be\label{eq:dmateig}
    \varrho = \sum_kp_k\ketbra{k}{k}.
\ee
For two arbitrary operators $A$ and $B,$ and a state $\varrho$ [Eq.~\eqref{eq:dmateig}], the QFI is defined as \cite{Paris2009,Braunstein1994,Holevo1982,Helstrom1976,
Petz2008}
\be
  \label{eq:fab}
  \qfif{\varrho,A,B}:=2\sum_{k, k'}
  \frac{(p_k-p_{k'})^2}{p_k+p_{k'}}
  {A}_{k,k'}{B}_{k',k},
\ee
where $A_{k,k'}=\braOket{k}{A}{k'}$ and $B_{k,k'}=\braOket{k}{B}{k'}$.
If the two operators are the same then, from Eq.~\eqref{eq:fab}, the usual form of the QFI is obtained:
\be
  \qfif{\varrho,A} \equiv \qfif{\varrho,A,A}=2\sum_{k, k'}
  \frac{(p_k-p_{k'})^2}{p_k+p_{k'}}
  |{A}_{k,k'}|^2. \label{eq:FQAA}
\ee

We list some useful properties of the QFI:
\pagebreak

(i) Based on Eq.~(\ref{eq:fab}), $\qfif{\varrho, A, B}$ is linear in the second and third arguments
\be
  \label{eq:qfi-linear-in-arguments}
  \qfif{\varrho,\sum_i A_i,\sum_j B_j} = \sum_{i,j} \qfif{\varrho,A_i,B_j}.
\ee
This will make it possible to calculate the QFI for
collective quantities based on the QFI for
single-particle observables.

(ii) The QFI remains invariant if we exchange the second and the third arguments
\be\label{eq:fab-fba}
    \qfif{\varrho,A,B}=\qfif{\varrho,B,A}.
\ee
Equation~\eqref{eq:fab-fba} will help to simplify our calculations.

(iii) The following alternative form,
\be\label{eq:fab-rewritten}
    \qfif{\varrho, A, B} = 4 \meanO{AB} - 8\sum_{k,k'} \frac{p_kp_{k'}}{p_k+p_{k'}} {A}_{k,k'}{B}_{k',k},
\ee
is also useful since the correlation appears explicitly.

(iv) For pure states, Eq.~\eqref{eq:fab} simplifies to
\be\label{eq:fab-pure-states}
    \qfif{\ket{\psi},A,B}=4\left(\meanO{AB}_{\psi} -\meanO{A}_{\psi}\meanO{B}_{\psi}\rpar.
\ee
Using Eq.~\eqref{eq:fab-pure-states} for $A{=}B,$ we obtain that for pure states the QFI equals four times the variance, i.e., $\qfif{\ket{\psi},A}=4(\Delta A)^2.$

(v) The QFI is convex on the space of the density matrices, i.e.,
\be
  \qfif{p\varrho_1{+}(1{-}p)\varrho_2,A}\leqslant
  p\qfif{\varrho_1,A}{+}(1{-}p)\qfif{\varrho_2,A},
\ee
Hence, when maximizing the QFI,
 we need to carry out an optimization over pure states only.

In the following, we show the general form of the expressions giving the precision bounds for states insensitive to the homogeneous field, as well as for states sensitive to it.
We also show that both bounds are invariant under the spatial translation of the system which makes the computing for particular cases much easier.

\subsection{Precision bound  for states insensitive to homogeneous fields:
Single-parameter dependence}

We will now consider quantum states that are  insensitive to the homogeneous field. For such states,
\be[\varrho, H_0]=0\label{eq:comm_insensitive}\ee
holds.
Hence, the unitary time evolution given in Eq.~\eqref{eq:unitary-operator} is simplified to
\be
U=e^{-ib_1H_1},\label{eq:Utrans_insensitive}
\ee
and the evolved state is a function of a single unknown parameter $b_1.$

When estimating a single parameter, the Cram\'er-Rao bound gives the best achievable precision as \cite{Paris2009}
\be
  \label{eq:one-parameter-precision}
  \varinv{b_1}|_{\max} = \qfif{\varrho,H_1}.
\ee
It is always possible to find a measurement that saturates the precision bound, \eqref{eq:one-parameter-precision}, which is indicated using the notation ``$|_{\max} = $''.

\pagebreak
\begin{observ}
  \label{obs:bound-for-insensitive-and-thermal-state}
  For states insensitive to the homogeneous fields, the maximal precision of the estimation of the gradient parameter $b_1$ is given as
  \be
    \label{eq:bound-for-insensitive-and-thermal-state}
    \varinv{b_1}|_{\max} = \sum_{n,m}^N \int x_n x_m P(\bs{x}) \dif{\bs{x}}\, \qfif{\varrho^{(\text{s})}, j_z^{(n)}, j_z^{(m)}},
  \ee
  where the integral represents the correlation between the particle positions $x_n$ and $x_m$.
  Moreover, Eq.~\eqref{eq:bound-for-insensitive-and-thermal-state} is translationally invariant, i.e., it remains the same after an arbitrary displacement $d$ of the form of
  \be
    U_d=\exp(-idP_x),
    \label{eq:unitary-translation}
  \ee
where $d$ is the distance displaced and $P_x$ is the sum of all single-body momentum operators $p_x^{(n)}$ in the $x$ direction.
\end{observ}

\begin{proo} We have to evaluate the right-hand side of Eq.~\eqref{eq:one-parameter-precision}.
The state is a tensor product of the spatial and internal parts, and the spatial part is an incoherent mixture of position eigenstates, as in Eqs.~\eqref{eq:state-internal-external} and \eqref{eq:thermal-state}.
Hence, the eigenstates are
$\ket{\bs{x},\lambda}$, where $\ket{\bs{x}}$ and $\ket{\lambda}$ are defined in the spatial and internal Hilbert spaces, respectively.
Then, the matrix elements of $H_1$, which is diagonal in the spatial subspace, are obtained as
\be
  \label{eq:generator1-diagonal-spatially}
  \begin{split}
    (H_1)_{\bs{x},\lambda;\bs{y},\nu}
    &=
    \delta(\bs{x}-\bs{y}) \braOket{\lambda}{\sum_{n=1}^N x_n j^{(n)}}{\nu}.
  \end{split}
\ee
Calculating Eq.~\eqref{eq:FQAA} for $A\,{=}\,H_1$, Eq.~\eqref{eq:one-parameter-precision} leads to Eq.~\eqref{eq:bound-for-insensitive-and-thermal-state} (see Appendix~\ref{ap:long-calc} for details).

In the last part of the proof, we show that  the precision \eqref{eq:bound-for-insensitive-and-thermal-state} remains the same for any displacement of the system. We use the Heisenberg picture in which the operators must be transformed instead of the states.
After the displacement, the operator $H_1$ describing the effect of the gradient  is obtained as
\be
\begin{split}
    \label{eq:shifted h1 generator}
    H_1(d)
     &=H_1-dH_0.
\end{split}
\ee
Hence, the unitary evolution operator of the displaced system is obtained as
\be\label{eq:Utrans}
    U(d)=e^{-i[b_0H_0+b_1 H_1(d)]}=e^{-i[(b_0-b_1d)H_0+b_1H_1]}.
\ee
Using the commutation relation \eqref{eq:comm_insensitive}, we can see that  Eq.~\eqref{eq:Utrans} is equal to the time evolution given in Eq.~\eqref{eq:Utrans_insensitive}.
\end{proo}

\subsection{Precision bound for states sensitive to homogeneous fields:
Two-parameter dependence}

\label{sec:Precision bound for states sensitive to homogeneous fields: Two-parameter dependence}

We now show how to obtain the precision bounds for states sensitive to the
homogeneous field. The homogeneous field rotates all the spins in the same way,
while the field gradient rotates the spins
differently depending on the position of the particles.
Hence, in order to estimate $b_1,$ we have to consider the effect of a second
unknown parameter $b_0$. Note, however, that we are not interested to estimate $b_0$ precisely, we just need it to estimate $b_1.$

In this case, the metrological performance of the quantum state is given by the $2\times2$ Cram\'er-Rao matrix inequality \cite{Paris2009}
\be
\bs{C} \ge \bs{\mathcal{F}}_{\text{Q}}^{-1},\label{eq:CR_matrix}
\ee
where the covariance matrix is defined as $C_{ij}=\ex{b_i b_j} - \ex{b_i}\ex{b_j}$.
The matrix elements of the quantum Fisher\break information matrix $\bs{\mathcal{F}}_{\text{Q}}$ are
\be
  \mathcal{F}_{ij}:= \qfif{\varrho, H_i, H_j}.
\ee
Unlike in the case of single-parameter estimation, Eq.~\eqref{eq:CR_matrix} can be saturated  only if the measurements for estimating the two parameters are compatible with each other \cite{Helstrom1976,
Paris2009, Ragy2016}.
Hence, we use ``$\leqslant$'' instead of ``$|_{\max}$'' for the bounds for quantum states sensitive to the homogeneous fields.

Using the well-known formula for the inverse of $2\times 2$ matrices,  Eq.~\eqref{eq:CR_matrix} yields \be
    \label{eq:precision bound for b1 in terms of QFI matrix elements}
    \varinv{b_1}\leqslant \mathcal{F}_{11}-\frac{\mathcal{F}_{01}\mathcal{F}_{10}}{\mathcal{F}_{00}},
\ee
for the precision of $b_1$.

\begin{observ}
\label{obs:bound-for-sensitive-and-thermal-state}
For states sensitive to the homogeneous field, the expression to compute the precision bound for the gradient parameter takes the following form:
    \be
    \label{eq:bound-for-sensitive-and-thermal-state}
    \begin{split}
        \varinv{b_1} \leqslant& \sum_{n,m}^N \int x_nx_m P(\bs{x})\dif{\bs{x}} \qfif{\varrho^{(\text{s})}, j_z^{(n)}, j_z^{(m)}}\\
         -& \frac{\lpar\sum_{n=1}^N  \int x_n P(\bs{x}) \dif{\bs{x}} \qfif{\varrho^{(\text{s})}, j_z^{(n)},J_z} \rpar^2}{\qfif{\varrho^{(\text{s})}, J_z}}.
    \end{split}
    \ee
    Moreover, the bound, \eqref{eq:bound-for-sensitive-and-thermal-state}, similarly to Eq.~(\ref{eq:bound-for-insensitive-and-thermal-state}), is invariant under spatial translations of the system.
\end{observ}

\begin{proo}
  To obtain the bound, (\ref{eq:bound-for-sensitive-and-thermal-state}), we need to consider the matrix elements of QFI one by one.
  First of all, we compute $\mathcal{F}_{11}$ which has the same form as Eq.~(\ref{eq:bound-for-insensitive-and-thermal-state})
  \be\label{eq:f11}
    \mathcal{F}_{11}=\sum_{n,m}^N \int x_n x_m P(\bs{x}) \dif{\bs{x}}\qfif{\varrho^{(\text{s})}, j_z^{(n)}, j_z^{(m)}}.
  \ee
  Next, we have that $H_0$, similarly to Eq.~\eqref{eq:generator1-diagonal-spatially}, is diagonal in the spatial $\ket{x}$ basis, and its matrix elements in the $\ket{\bs{x},\lambda}$ basis of the state are written as
  \be\label{eq:generator0-diagonal-spatially}
  \begin{split}
    (H_0)_{\bs{x},\lambda;\bs{y},\nu}
    =& \delta(\bs{x}-\bs{y})\braOket{\lambda}{\sum_{n=1}^N j_z^{(n)}}{\nu}.
  \end{split}
  \ee
  With this we obtain $\qfif{\varrho, H_0, H_0}$ as
  \be\label{eq:f00}
    \mathcal{F}_{00} = \qfif{\varrho^{(\text{s})}, J_z}.
  \ee
  Note that Eq.~\eqref{eq:f00} is not a function of the whole state but only of the internal $\varrho^{(\text{s})}$ state.
  Finally, we compute $\mathcal{F}_{01}$ and $\mathcal{F}_{10}$.
  Since $\mathcal{F}_{01}\,{=}\,\mathcal{F}_{10}$, we have to compute only one of them.
  Using Eqs.~\eqref{eq:generator0-diagonal-spatially} and \eqref{eq:generator1-diagonal-spatially},
   $\qfif{\varrho,H_0,H_1}$ is obtained as
  \be\label{eq:f01}
    \mathcal{F}_{01} = \sum_{n=1}^N \int x_n P(\bs{x})\dif{\bs{x}} \qfif{\varrho^{(\text{s})}, j_z^{(n)},J_z}.
  \ee
  With these results, Eq.~\eqref{eq:bound-for-sensitive-and-thermal-state} follows (see Appendix~\ref{ap:long-calc}).

Let us now determine the bound on the precision for estimating the gradient on the translated system. We have to compute first the QFI matrix elements.
We use the linearity of the last two arguments of $\qfif{\varrho, A, B}$ given in Eq.~(\ref{eq:qfi-linear-in-arguments}), the fact that $H_0$ remains unchanged in the Heisenberg picture. We also use the formula (\ref{eq:shifted h1 generator}) for the shifted  $H_1$ operator.
The diagonal element of the QFI matrix corresponding to the measurement of the homogeneous field is
  \begin{align}
      \mathcal{F}_{00}(d)&=\qfif{\varrho, H_0(d)} = \mathcal{F}_{00},
      \label{eq:shifted h1 generator F00}
  \end{align}
 hence, it does not change due to the translation.
For the diagonal element corresponding to the gradient measurement we obtain
     \begin{align}
      \mathcal{F}_{11}(d)
        &=\mathcal{F}_{11}-2d\mathcal{F}_{01}+d^2\mathcal{F}_{00}.
        \label{eq:shifted h1 generator F11}
  \end{align}
 Finally, for the off-diagonal element, we get
  \begin{align}
      \mathcal{F}_{01}(d)
        &=
        \mathcal{F}_{01}-d\mathcal{F}_{00}.
        \label{eq:shifted h1 generator F01}
  \end{align}
  After determining all the elements of the QFI matrix, the bound for a displaced system can be obtained as
  \be
  \begin{split}
      \varinv{b_1}\leqslant\,&\mathcal{F}_{11}(d) -\frac{(\mathcal{F}_{01}(d))^2}{\mathcal{F}_{00}(d)}\\
      =\,& \mathcal{F}_{11}-2d\mathcal{F}_{01}+d^2\mathcal{F}_{00}\\
      &-\frac{\mathcal{F}_{01}^2-2d \mathcal{F}_{01}\mathcal{F}_{00} +d^2\mathcal{F}_{00}^2}{\mathcal{F}_{00}}.
      \label{eq:varb_sensitive}
  \end{split}
  \ee
  The bound in Eq.~\eqref{eq:precision bound for b1 in terms of QFI matrix elements} can be obtained from the right-hand side of Eq.~\eqref{eq:varb_sensitive} with straightforward algebra.
\end{proo}


\section{Spin chain and two separated ensembles for magnetometry}
\label{sec:twin cloud systems}

After presenting our tools in Sec.~\ref{sec:cramer-rao bounds}, we start with simple examples to show how our method works.
We calculate precision bounds for gradient metrology for spin chain and for two-particle ensembles separated by a distance.

Before considering the setups mentioned above,
we introduce various quantities describing the distribution of the particles based on the probability distribution function  appearing in Eq.~\eqref{eq:thermal-state}.
The mean particle position is
\be\label{eq:mean}
    \mu = \int \frac{\sum_{n=1}^N x_n}{N} P(\bs{x}) \dif{\bs{x}}.
\ee
The standard deviation of the particle positions, describing the size of the system, is computed as
\be\label{eq:variance}
    \sigma^2 = \int \frac{\sum_{n=1}^N x_n^2}{N} P(\bs{x}) \dif{\bs{x}} - \mu^2.
\ee
Finally, the covariance averaged over all particle pairs is
\be\label{eq:covariance}
    \eta = \int \frac{\sum_{n\neq m}^N x_nx_m}{N(N-1)} P(\bs{x}) \dif{\bs{x}} - \mu^2.
\ee
The covariance is a large positive value if the particles tend to be close to each other,
while it is negative if they tend to avoid each other.

\begin{figure}[htp]
\begin{center}
\includegraphics[width=210pt]{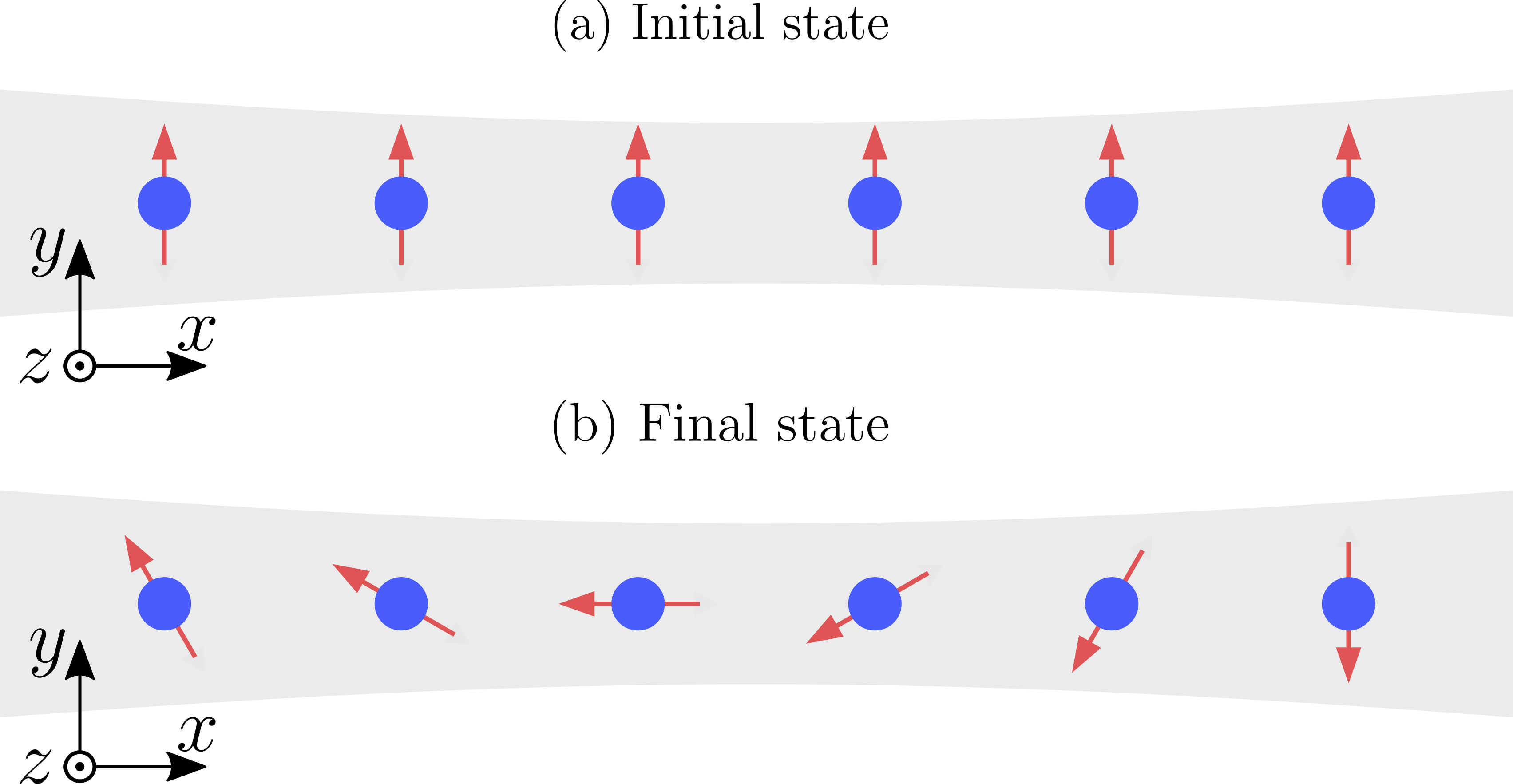}
\caption{
    A one-dimensional chain of six spin-$j$ atoms (blue
    disks) confined in a potential (gray area).
    (a) The ensemble is initially totally polarized along the $y$ direction.
    The magnetic field $B_z$ points outward from the figure.
    The spin chain is along the $x$ direction.
    (b) If the magnetic field has a nonzero gradient, then it affects the spins of the individual atoms differently depending on the position of the atoms.\label{fig:ionchain-evolution}}
\end{center}
\end{figure}

After presenting the fundamental quantities above, let us study concrete metrological setups.
The first spatial state we consider is given by $N$
particles placed equidistantly from each other in a one-dimensional spin chain, as
shown in Fig.~\ref{fig:ionchain-evolution}.
Such a system has been studied also in the context of a single parameter estimation in the presence of collective phase noise \cite{Altenburg2016}.
The probability density function describing such a system is
\be\label{eq:pdf-spin-chain}
  P(\bs{x})=\prod_{n=1}^N \delta(x_n-na),
\ee
where $a$ is the distance between the particles in the chain.
For this system, the average position of the $n{\text{th}}$ particle is
\be\label{eq:mean-chain}
  \int x_n P(\bs{x}) \dif{\bs{x}}= na,
\ee
whereas the two-point average \eqref{eq:covariance} is
\be\label{eq:second-moment-chain}
  \int x_nx_m P(\bs{x}) \dif{\bs{x}}= nma^2.
\ee
The standard deviation defined in Eq.~\eqref{eq:variance} is obtained as
\be
    \label{eq:variance-chain}
    \sigma_{\text{ch}}^2 = a^2\frac{N^2-1}{12}.
\ee

Next, we will obtain precision bounds for particles placed in a spin chain.

\begin{observ}
    Let us consider a chain of $N$ spin-$j$ particles placed along the $x$ direction separated by a constant distance, and a magnetic field pointing in the $z$ direction.
    Then, for the spin-state totally polarized in the $y$ direction,
 \be\label{eq:totally polarized state}
    \ket{\psi_{\text{tp}}}=\ket{j}_y^{\otimes N},
\ee
the precision bound is given by
    \be\label{eq:precision-tp-ch}
        \varinv{b_1}|_{\max} = 2\sigma_{\text{ch}}^2Nj.
    \ee
    Here, $\sigma_{\text{ch}}$ denotes the standard deviation of the average position of the particles for the chain (ch).
\end{observ}

\begin{proo}
  We use the precisions bound for states sensitive to the homogeneous field given in Eq.~\eqref{eq:bound-for-sensitive-and-thermal-state}. We obtain
\be
\label{eq:result-for-chain-without-sigma}
\begin{split}
  \varinv{b_1}|_{\max} &= \, \sum_{n,m}^N nma^2 \qfif{\ket{j}_y^{\otimes N}, j_z^{(n)},j_z^{(m)}}\\
  &-\frac{\lpar \sum_{n=1}^N an \qfif{\ket{j}_y^{\otimes N}, j_z^{(n)}, J_z} \rpar^2}{\qfif{\ket{j}_y^{\otimes N}, J_z, J_z}}\\
  &=\, 2a^2 \frac{N^2 -1}{12} Nj,
\end{split}
\ee
Note that the bound can be saturated (see Appendix~\ref{app:compatibility-of-measurements}).
Here, for the last equality we used the definitions of the average quantites given in  Eqs.~\eqref{eq:mean-chain} and \eqref{eq:second-moment-chain} and we also used Eq.~\eqref{eq:fab-pure-states} giving the QFI for pure states.
We can see that the standard deviation given in Eq.~(\ref{eq:variance-chain}) coincides with a factor we have in Eq.~(\ref{eq:result-for-chain-without-sigma}), with which we conclude the proof.
\end{proo}

Note that the bound \eqref{eq:result-for-chain-without-sigma} seems to scale with the third power of the particle number $N$, and hence seems to overcome the ultimate Heisenberg limit.
The reason is that the length of the chain increases as we introduce more particles into the system. We should compare the metrological usefulness of systems with different particle numbers, but of the same size.
In our case, we  use throughout the paper the standard deviation of the averaged particle positions as a measure of the spatial size of the system, and normalize the results with it. One can miss this important point since when only the homogeneous field is measured such a normalization is not needed.\footnote{This comment is relevant for the setup of Ref.~\cite{Zhang2014}, where
the precision of the gradient estimation seems to reach the Heisenberg scaling. In reality, the shot-noise scaling has not been overcome. The question of normalization is also important for the setup in Ref.~\cite{Ng2014}.}

After the spin chain, we consider estimating the gradient with two
ensembles of spin-$j$ atoms spatially separated from each other. Such systems have been realized in cold gases (e.g., Ref.~\cite{Julsgaard2001}), and can be used for differential interferometry \cite{Eckert2006, Landini2014, Altenburg2016}.
We will determine the internal state with the maximal QFI.

Let us assume that
half of the particles are at one position and the rest at another one, both places at a distance $a$  from the origin.
The probability density function of the spatial part is
\be
  \label{eq:double-well-spatial-pdf}
  P(\bs{x})=\prod_{n=1}^{N/2} \delta(x_n+ a)\prod_{n=N/2+1}^{N} \delta(x_n-a).
\ee
Such a distribution of particles could be realized in a double-well trap, where the width of the wells is negligible compared to the distance between the wells.
To distinguish the two wells we use the labels ``L'' and ``R'' for the left-hand side and right-hand side wells, respectively.
Based on these, we obtain the single-point averages as
\begin{align}
  \label{eq:single-point-function-double-well}
  \int x_n P(\bs{x}) \dif{\bs{x}}&= \lcor\ver{-a \quad \text{if } n\in \text{L},}{+a \quad \text{if } n\in \text{R}.}\right.
  \end{align}
The two-point correlation functions are
  \begin{align}
  \label{eq:two-point-function-double-well}
  \int x_nx_m P(\bs{x}) \dif{\bs{x}}&= \lcor\ver{+a^2 \quad \text{if } (n,m)\in \text{(L,L) or (R,R),}}{-a^2 \quad \text{if } (n,m)\in \text{(L,R) or (L,R).}} \right.
\end{align}
For the average particle position we obtain $\mu=0,$ while the standard deviation for the spatial state in the double well (dw) is
  \be
    \sigma_{\text{dw}}^2 = a^2,
  \ee

Next, we calculate the achievable precision of the gradient estimation.

\begin{observ}
\label{obs:GHZ} For the case of two ensemble of $N$ spin-$j$ particles,
    the state that  maximizes the QFI is
        \be\label{eq:best-state}
        \ket{\psi} = \frac{\ket{j\cdots j}^{(\text{L})}\ket{{-}j\cdots {-}j}^{(\text{R})}+\ket{{-}j\cdots{-}j}^{(\text{L})}\ket{j\cdots j}^{(\text{R})}}{\sqrt{2}}.
    \ee
    \pagebreak

    \noindent
    The best achievable precision is given as
    \be\label{eq:bound-dw}
        \varinv{b_1}|_{\max} = 4 \sigma_{\text{dw}}^2 N^2j^2.
    \ee
    Equation~\eqref{eq:bound-dw} agrees with the results obtained in Ref.~\cite{Landini2014}.
\end{observ}

\begin{proo}
  The state given in Eq.~\eqref{eq:best-state} is insensitive to the homogeneous field, hence we have to use the formula \eqref{eq:bound-for-insensitive-and-thermal-state} to bound the precision. We obtain
  \be\label{eq:f11_dowble_well}
  \begin{split}
    \varinv{b_1}|_{\max} &= \sum_{\ver{(n,m)=}{\text{(L,L)}, \text{(R,R)}}} a^2 \qfif{\ket{\psi}, j_z^{(n)}, j_z^{(m)}} \\
    &+ \sum_{\ver{(n,m)=}{\text{(L,R)},\text{(R,L)}}}{-}a^2 \qfif{\ket{\psi}, j_z^{(n)}, j_z^{(m)}}.\\
    \end{split}
    \ee
    For the state \eqref{eq:best-state}, the equation above, \eqref{eq:f11_dowble_well}, yields
    \be
    \begin{split}
    \varinv{b_1}|_{\max}&= \sum_{\ver{(n,m)=}{ \text{(L,L)},   \text{(R,R)}}} a^2 j^2 + \sum_{\ver{(n,m)=}{ \text{(L,R)}, \text{(R,L)}}} -a^2 (-j^2)\\
    &= 4 a^2 N^2j^2,
  \end{split} \label{eq:dwell}
  \ee
  where we have used the definition of the QFI for pure states given in Eq.~\eqref{eq:fab-pure-states}. A factor in Eq.~\eqref{eq:dwell} can be identified with the standard deviation, \eqref{eq:variance-chain}, from which the proof follows.\end{proo}

It is interesting to simplify the QFI for product states states  $\ket{\psi}^{(\text{L})} \otimes \ket{\psi}^{(\text{R})}$, where $\ket{\psi}^{(\text{L})}$ and  $\ket{\psi}^{(\text{R})}$ are pure states of $N/2$ particles each.
This approach is also discussed in Ref.~\cite{Landini2014}.
Such states can reach the Heisenberg limit, while they are easier to realize experimentally than states in which the particles in the wells are entangled with each other.

Before obtaining the precision for the case above, we present a method to simplify our calculations.
The system is at the origin of the coordinate system such that for mean particle position given in Eq.~\eqref{eq:mean},
\be\label{eq:mu_rewritten}
    \mu=\int \frac{\sum_n x_n}{N} P(\bs{x})\dif{\bs{x}} = 0
\ee
holds. Thus, the second term in the expression for the bound for states sensitive to the homogeneous field \eqref{eq:bound-for-sensitive-and-thermal-state} is zero since all $\qfi[\varrho^{(\text{s})}, j_z^{(n)}, J_z]$ are equal considering product states of two equal permutationally invariant states, $\ket{\psi}^{(\text{L})} \otimes \ket{\psi}^{(\text{R})}$.
Hence, the bounds for states insensitive and sensitive to the homogeneous field, Eqs.~\eqref{eq:bound-for-insensitive-and-thermal-state} and \eqref{eq:bound-for-sensitive-and-thermal-state}, respectively, are the same in this case.

We now compute $\qfi[\rho, H_1]$ for the case when the state is sensitive to the homogeneous field,
hence, we use the bound on the precision given in Eq.~\eqref{eq:bound-for-sensitive-and-thermal-state}.
Using the the probability density distribution function given in Eq.~\eqref{eq:double-well-spatial-pdf}, and following steps leading to Eq.~\eqref{eq:dwell}, we obtain
\begin{equation}\label{eq:bound-product-state-dw}
    \qfif{\ket{\psi}^{(\text{L})}\ket{\psi}^{(\text{R})}, H_1}
    = 2a^2\qfif{\ket{\psi}^{(\text{L})},J_z^{(\text{L})}},
\end{equation}
where we used that $\qfif{\ket{\psi}^{(\text{L})},J_z^{(\text{L})}} = \qfif{\ket{\psi}^{(\text{R})},J_z^{(\text{R})}}$. Note that our results concerning using product states for magnetometry can be interpreted as follows. In this case, essentially the homogeneous field is estimated in each of the two wells, and then the gradient is computed from the measurement results.
The bounds for these type of states are also saturable (see Appendix~\ref{app:compatibility-of-measurements}).

We will now present precision bounds for various well known quantum states in the two wells.
We consider the Greenberger-Horne-Zeilinger (GHZ) state \cite{Greenberger1989,Pan2000, Yao2012, Lu2007,Sackett2000, Monz2011}
\be\label{eq:ghz-def}
    \ket{\ghz} = \frac{\ket{00\cdots00}+\ket{11\cdots11}}{\sqrt{2}},
\ee
where $\ket{0}$ and $\ket{1}$ are the eigenstates of $j_z^{(n)}$ with eigenvalues $-\frac{1}{2}$ and $+\frac{1}{2},$ respectively.
We also consider unpolarized Dicke states \cite{Dicke1954,Toth2007,Kiesel2007,Wieczorek2009,Chiuri2012,Luecke2011,Hamley2012,Haeffner2005}
\be\label{eq:dicke-def}
    \ket{\dicke{N}}_l=\binom{N}{N/2}^{-1/2}\sum_{k} {\mathcal{P}}_{k}(\ket{0}_l^{\otimes N/2} \otimes \ket{1}_l^{\otimes N/2}),
\ee
where $l=x,y,z$ and the summation is over all ${\mathcal{P}}_k$ permutations. Such states are the symmetric superposition of product states with an equal number of $\ket{0}_l$'s and $\ket{1}$'s.
Based on these, in Table~\ref{tab:result-states-two-ensembles} we summarized the precision bounds for states of the type $\ket{\psi}^{(\text{L})}\otimes \ket{\psi}^{(\text{R})}$ for the double-well case.

\begin{table}
  \caption{Precision for differential magnetometry for various product states of the type $\ket{\psi}^{\rm (L) }\otimes\ket{\psi}^{\rm (R) }$ in the two ensembles.
  Note that there are $N_{\text{L}}=N_{\text{R}}=N/2$ particles in each ensemble.
  In the second column we show the QFI for the estimation of the homogeneous field appearing in the literature, for states with $N_{\text{L}}$ particles.
  The third column shows the result for the bounds obtained with Eq.~\eqref{eq:bound-product-state-dw}.}
  \label{tab:result-states-two-ensembles}
  \renewcommand\arraystretch{1.4}
  \begin{tabular}{
    m{0.12\linewidth}
    >{\centering}m{0.57\linewidth}
    >{\centering\arraybackslash}m{0.25\linewidth} }
    \hline
    \hline
    $\ket{\psi}$ & $\qfif{\ket{\psi},J_z}$ & $\varinv{b_1}|_{\max}$ \\[0.2em]
    \hline
    $\ket{j}_{y}^{\otimes N_{\text{L}}}  $ & $2N_{\text{L}}j$ & $2a^2Nj$ \\
    $\ket{\psi_{\text{sep}}} $ & $ 4N_{\text{L}}j^2$ & $ 4a^2Nj^2$ \\
    $\ket{\ghz} $ & $N_{\text{L}}^2$ & $a^2N^2/2 $\\
    $\ket{\dicke{N_{\text{L}}}}_x$ & $N_{\text{L}}(N_{\text{L}}+2)/2$ & $a^2N(N+4)/4$\\
    \hline
    \hline
  \end{tabular}
\end{table}

\vspace{-1em}

\section{Magnetometry with a single\\ atomic ensemble}
\label{sec:single cloud systems}
\vspace{-.5em}

\begin{figure}[!b]
\begin{center}
\includegraphics[width=210pt]{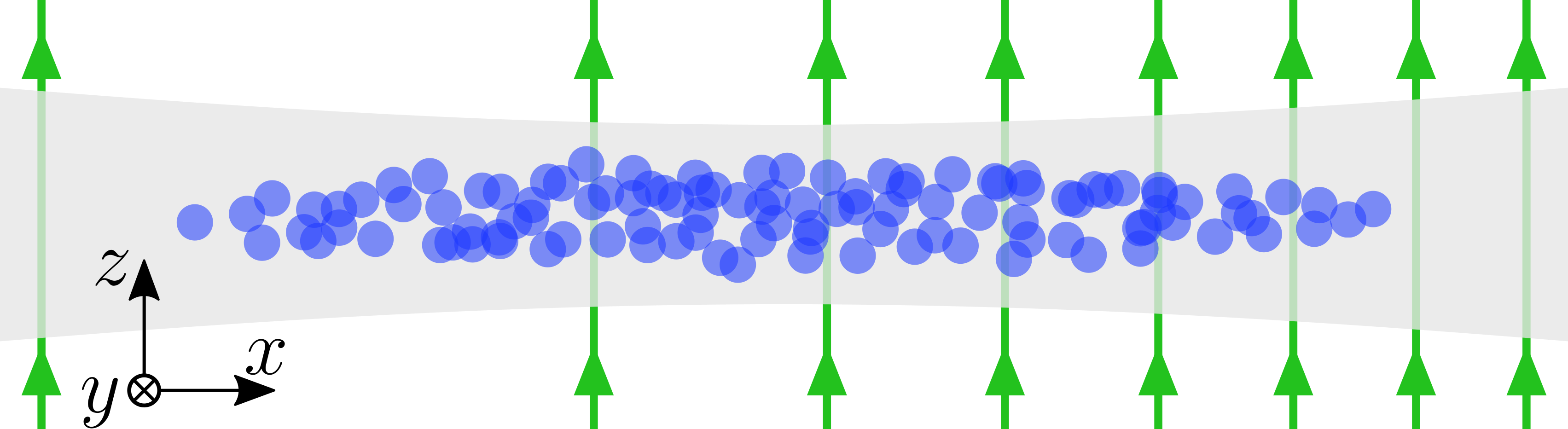}
\caption{
An ensemble of spin-$j$ atoms in a cigar-shaped trap elongated in the $x$ direction.
The magnetic field, represented by green arrows,  points in the $z$ direction and it is linear in $x$. Its strength is proportional to the density of the field lines.
}
\label{fig:single cloud particle density under the magnetic gradient field}
\end{center}
\end{figure}

In this section, we discuss magnetometry with a single
atomic ensemble.
We consider a one-dimensional ensemble of spin-$j$ atoms
placed in a trap which is elongated
in the  $x$ direction.
The setup is depicted
in Fig.~\ref{fig:single cloud particle density under the magnetic gradient field}.
In the second part of the section, we calculate precision bounds for the
gradient estimation with some important multiparticle quantum states,
for instance, Dicke states, singlet states, and GHZ states.

\vspace{-0.5em}
\subsection{Precision bound for an atomic ensemble}
\vspace{-0.5em}

In an atomic ensemble of many atoms, typically
the atoms cannot be individually addressed.
We will  take this into account by considering states for which both the internal state $\varrho^{(\text{s})}$ and the probability distribution function
$P(\bs{x})$, appearing in Eq.~(\ref{eq:thermal-state}),
are PI.
The permutational invariance of $P(\bs{x})$ implies that
\be
  \label{eq:pi-for-pdf}
  P(\bs{x})=\tfrac{1}{N!}\sum_{k}\mathcal{P}_k [P(\bs{x})]
\ee
holds, where the summation is over all possible permutations $\mathcal{P}_k$
of the variables $x_n$.
Hence, we do not need to sum over all possible $n$'s  in Eqs.~\eqref{eq:mean} and \eqref{eq:variance}, and neither to sum over all $n$'s and $m$'s in Eq.~\eqref{eq:covariance}.
All the terms in each sum are equal to each other due to the permutationally invariance of the probability distribution function \eqref{eq:pi-for-pdf}.

\begin{figure}[tp]
\begin{center}
    \vskip0.5cm\includegraphics[width=210pt]{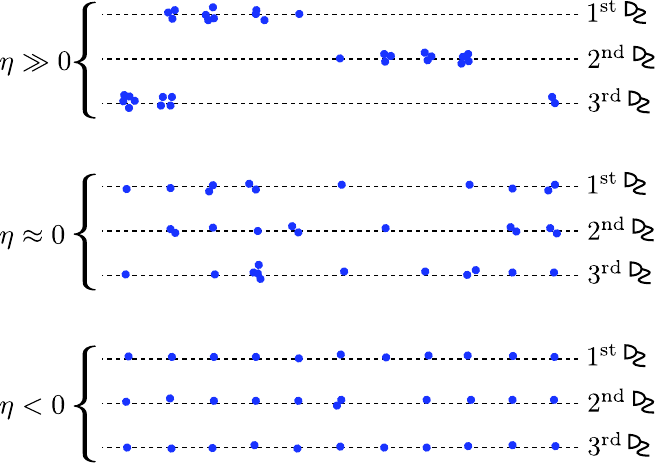}\vskip0.35cm
    \caption{Possible particle distributions for a chain of atoms on a 1D lattice,
    assuming that Eq.~\eqref{eq:pi-for-pdf} holds for the probability distribution function.
    We consider three cases, determined by the value of the covariance $\eta$.
        (Top) The atoms bunch together due to high correlation in the positions. As explained in the text, this leads to the possibility of estimating well the magnetic field at each point, and hence possibly obtaining a very good estimate of the gradient parameter.
    (Middle) Due to the small covariance the
    particles fill up the sites more uniformly.
    (Bottom) The negative correlation makes the particles be far from
    each other, filling the trap uniformly.
    }
    \label{fig:covariance examples}
\end{center}
\end{figure}

An interesting property of the covariance \eqref{eq:covariance} is that it can only take values bounded by the variance in the following way:
\be\label{eq:etabounds}
  \frac{-\sigma^2}{N-1}\leqslant \eta\leqslant \sigma^2,
\ee
where both the lower and the upper bounds are proportional to the variance $\sigma^2.$
See Fig.~\ref{fig:covariance examples} for examples on how different correlations are obtained in an atomic 1D lattice.

Next, we present precision bounds for PI states.

\begin{observ}
  \label{obs:bound-insensitive-single-ensemble}
  The maximal precision achievable by a single atomic ensemble
  insensitive to homogeneous fields is
  \bea
  \label{eq:precision bound for single-cloud systems IH}
  \varinv{b_1}|_{\max} &=& (\sigma^2-\eta) \sum_{n=1}^{N} \qfif{\varrho^{(\text{s})},j_z^{(n)}}.
  \eea
  The precision given in Eq.~\eqref{eq:precision bound for single-cloud systems IH}
  can be reached by an optimal measurement.
  Nevertheless, it is worth to note that the precision cannot surpass the
  shot-noise scaling because $\qfif{\varrho^{(\text{s})},j_z^{(n)}}$
  cannot be larger than $j^2.$ Moreover, $\eta$ cannot be smaller than $-\sigma^2/(N-1)$ due to Eq.~\eqref{eq:etabounds}, which makes its contribution negligible for large $N$.
\end{observ}

\begin{proo}
 From the definition of the QFI for states insensitive to the homogeneous field [Eq.~\eqref{eq:bound-for-insensitive-and-thermal-state}] we obtain the bound for a single ensemble as
  \be
  \begin{split}
    \varinv{b_1}|_{\max} &= \sum_{n,m}^N \int x_nx_m P(\bs{x}) \dif{\bs{x}}\qfif{\varrho^{(\text{s})}, j_z^{(n)}, j_z^{(m)}}\\
    &=\sum_{n=1}^N \sigma^2 \qfif{\varrho,j_z^{(n)}} + \sum_{n\neq m}^N \eta \qfif{\varrho,j_z^{(n)},j_z^{(m)}}.
  \end{split}\label{eq:max_insensitive}
  \ee
    Then, we have to use the fact that for states insensitive to the homogeneous fields $\qfif{\varrho,J_z}=0$ holds, which implies
  \be
    \qfif{\varrho, J_z} = \sum_{n,m}^N \qfif{\varrho,j_z^{(n)},j_z^{(m)}} = 0.
  \ee
  Based on this, for such states the sum of QFI terms involving two operators can be expressed with the sum of QFI terms involving a single operator as
  \be
    \sum_{n\neq m}^N \qfif{\varrho,j_z^{(n)},j_z^{(m)}}=-\sum_{n=1}^N \qfif{\varrho, j_z^{(n)}}.
    \label{eq:qfi-identity-insensitive}
  \ee
  Substituting Eq.~\eqref{eq:qfi-identity-insensitive} into Eq.~\eqref{eq:max_insensitive}, Observation~\ref{obs:bound-insensitive-single-ensemble} follows.
\end{proo}

\begin{observ}
  \label{obs:bound-sensitive-single-ensemble}
  For states sensitive to homogeneous fields,
  the precision of estimating the gradient is bounded from above as
  \be
    \label{eq:precision bound for single-cloud systems SH}
    \varinv{b_1}|_{\max} = (\sigma^2-\eta) \sum_{n=1}^N \qfif{\varrho^{(\text{s})},j_z^{(n)}}+\eta \qfif{\varrho^{(\text{s})},J_z},
  \ee
  which may surpass the shot-noise scaling whenever $\eta$ is a positive constant.
\end{observ}

\begin{proo}
We start from Eq.~\eqref{eq:bound-for-sensitive-and-thermal-state} and take into account that in this case the bound is saturable (see Appendix~\ref{app:compatibility-of-measurements}).
As explained in Sec.~\ref{sec:Precision bound for states sensitive to homogeneous fields: Two-parameter dependence}, if we move the system, the precision bounds do not change.
We then move our system  to the origin of the coordinate system yielding $\mu=0,$
and making the second term appearing in Eq.~\eqref{eq:bound-for-sensitive-and-thermal-state}  zero.
Thus, we only compute the first term in Eq.~\eqref{eq:bound-for-sensitive-and-thermal-state} and obtain
  \be
  \begin{split}
    \varinv{b_1}|_{\max} =&
    \sum_{n=1}^N \sigma^2 \qfif{\varrho,j_z^{(n)}} + \sum_{n\neq m}^N \eta \qfif{\varrho,j_z^{(n)},j_z^{(m)}},
  \end{split}
  \ee
  Then, we add $\eta\sum_{n=1}^N \qfif{\varrho,j_z^{(n)}}$ to the last term and subtract it from the first term to make the expression more similar to Eq.~(\ref{eq:precision bound for single-cloud systems IH}).
  \end{proo}

  Note
  that the second term on the right-hand side of
  Eq.~(\ref{eq:precision bound for single-cloud systems SH})
  is new in the sense that it did not appear in
  the bound for states insensitive to
  homogeneous fields given in  Eq.~(\ref{eq:precision bound for single-cloud systems IH}).
  Even if the first term cannot overcome the shot-noise limit, in the second term the covariance is multiplied by the QFI for estimating the homogeneous field and therefore this concrete term, for extremely correlated particle positions, allows to achieve the Heisenberg scaling.

\subsection{Precision limit for various spin states}

In this section, we present the precision limits for various classes of important quantum states such as the totally polarized state, the state having the best precision among separable states, the singlet state, the Dicke state \eqref{eq:dicke-def}, or the GHZ state \eqref{eq:ghz-def}.
We calculate the precision bounds presented before, \eqref{eq:precision bound for single-cloud systems IH} and \eqref{eq:precision bound for single-cloud systems SH}, for these systems.

\subsubsection{Singlet states}

A pure singlet state is a simultaneous eigenstate of the collective $J_z$ and $J^2$
operators, with an eigenvalue zero for both operators.
We will now consider PI singlet states.
Surprisingly, the precision bound is the same for any such state. PI singlet states are very relevant for experiments, since they have been experimentally
created in cold gases \cite{Toth2010,Behbood2014} while they also appear in condensed matter physics \cite{Jakab2018The}.

Let us now see the most important properties of singlet states of an $N$-particle system. There are several singlets pairwise orthogonal to each other. The number of such singlets, $D_0$, depends on the particle spin $j$ and the number of particles $N$.
It is the most natural to write the singlet state in the angular momentum basis. The basis states are $\ket{J,M_z,D},$
which are the eigenstates of $J_x^2+J_y^2+J_z^2$ with an eigenvalue $J,$ and of $J_z$ with an eigenvalue $M_z.$
The label $D$ is used to distinguish different eigenstates corresponding to the same eigenvalue of $J$ and $J_z$.
Then, a singlet state can be written as
\be
  \varrho_{\text{singlet}}^{(\text{s})}=\sum_{D=1}^{D_0} p_D\ketbra{0,0,D}{0,0,D},
  \label{eq:definition of a general singlet}
\ee
where $\sum_D p_D{=}1$.

Let us see some relevant single-particle expectation values for the singlet.
Due to the rotational invariance of the singlet $\varrho_{\text{singlet}}^{(\text{s})}$, we obtain that
\be
    \ex{(j_x^{(n)})^2}=\ex{(j_y^{(n)})^2}=\ex{(j_z^{(n)})^2}
\ee
holds.
We also know that for the sum of the second moments of the single particle angular momentum components
\be\label{eq:jx2jy2jz2}
\meanO{(j_x^{(n)})^2+(j_y^{(n)})^2+(j_z^{(n)})^2}=j(j+1)
\ee
holds.
Hence, the expectation value of the second moment of the single-particle angular momentum component is obtained as
\be\label{eq:trace of jzn square times the general singlet}
    \ex{(j_z^{(n)})^2}=\frac{j(j+1)}{3}.
\ee

After discussing the main properties of the singlet states, we can now obtain a precision bound for gradient metrology with such states.

\begin{observ}
  \label{obs:precision bound for the singlet state}
  For PI spin states living in the singlet subspace, i.e., states composed of vectors that have zero eigenvalues for $J_z$ and $J^2$ and all their possible statistical mixtures, the precision of the magnetic gradient parameter is bounded from above as
  \be
    \label{eq:sing_QFI}
    \varinv{b_1}_{\text{singlet}}|_{\max} = (\sigma^2-\eta) N \frac{4j(j+1)}{3}.
  \ee
\end{observ}

\begin{proo}
First compute the QFI for the one-particle operator $j_z^{(n)},$ $\qfif{\varrho^{(s)}, j_z^{(n)}}.$ For that we need that when $j_z^{(n)}$ acts on a singlet state, produces a state outside of the singlet subspace. Hence,
\be
\label{eq:jzn in subspace of singlets is a null operator}
\braOket{0,0,D}{j_z^{(n)}}{0,0,D'}=0
\ee
for any pair of pure singlet states.
Then, we use the formula \eqref{eq:fab-rewritten} to compute the QFI. The second term of Eq.~\eqref{eq:fab-rewritten} is obtained as
\be
    8 \sum_{D,D'} \frac{p_Dp_{D'}}{p_D+p_{D'}} |\braOket{0,0,D}{j_z^{(n)}}{0,0,D'}|^2 = 0,
\ee
due to Eq.~\eqref{eq:jzn in subspace of singlets is a null operator}.
It follows that the single-particle QFI for any singlet equals four times the second moment of the angular momentum component
\be
\label{eq:qfi to trace in the case of singlet}
\qfif{\varrho_{\text{singlet}}^{(\text{s})}, j_z^{(n)}}
=4\tr[{\varrho_{\text{singlet}}^{(\text{s})} (j_z^{(n)})^2}].
\ee
Note that Eq.~\eqref{eq:qfi to trace in the case of singlet} is true even though $\varrho_{\text{singlet}}^{(\text{s})}$ is a mixed state.
Inserting the expectation value of the second moment of the angular momentum component given in Eq.~\eqref{eq:trace of jzn square times the general singlet} into Eq.~\eqref{eq:qfi to trace in the case of singlet}, we
obtain $\qfif{\varrho_{\text{singlet}}^{(\text{s})}, j_z^{(n)}}$ for any $n.$
Then, we have all the ingredients to evaluate the maximal precision
given in Eq.~\eqref{eq:precision bound for single-cloud systems IH}, and with that we prove the Observation.
\end{proo}

As mentioned earlier, singlet states are insensitive
to homogeneous magnetic fields,
hence determining the gradient leads to a single-parameter
estimation problem.
This implies that there is an optimal operator that saturates the precision
bound given by Eq.~\eqref{eq:sing_QFI}.
However, it is usually very hard to find
this optimal measurement,
although a formal procedure for this exists \cite{Paris2009}.
In Ref.~\cite{Urizar-Lanz2013}, a particular setup for determining the magnetic gradient
with PI singlet states was suggested by the measurement
of the $J_x^2$ collective operator.
 For this scenario the precision is given by
\be
\label{eq: Jx2_acc}
\varinv{b_1}
= \frac{|\partial_{b_1}\ex{J_x^2}|^2}{\varian{J_x^2}}.
\ee
In Appendix~\ref{app: Optimal measurements for singlet states},
we  show that this measurement provides an optimal precision for gradient metrology for all PI singlets.

\subsubsection{Totally polarized state}
\label{sec:Totally polarized state}

The totally polarized state can easily be prepared experimentally.
It has already been used for gradient magnetometry with a single atomic ensemble \cite{Koschorreck2011,Vengalattore2007}.
For the gradient measurement as for the measurement of the homogeneous field, the polarization must be perpendicular to the field we want to measure.

We chose as before the totally polarized state along $y$ axis, given in Eq.~\eqref{eq:totally polarized state}.
The relevant variances for the state, \eqref{eq:totally polarized state}, are
\begin{subequations}
\bea\label{eq:totally polarized state, relevant variances}
(\Delta J_z)^2_{\rm tp}&=&Nj/2,\\\label{eq:totally polarized state, relevant variances b}
(\Delta j_z^{(n)})_{\rm tp}^2&=&j/2,
\eea
\end{subequations}
for all $n$.
Based on Eq.~\eqref{eq:fab-pure-states}, for pure states the QFI is just four times the variance. Hence, from Eq.~\eqref{eq:totally polarized state, relevant variances b}, we obtain $\qfif{\varrho,j_z^{(n)}}=2j$ and $\qfif{\varrho,J_z}=2Nj$.
Then, the bound on the sensitivity can be obtained from the formula for PI states sensitive to homogeneous fields (\ref{eq:precision bound for single-cloud systems SH}) as
\be\label{eq:precision bound for the totally polarized state}
\varinv{b_1}_{\text{tp}}|_{\max} = 2 \sigma^2 Nj.
\ee
We can see clearly that the precision scales as $\mathcal{O}(N)$ for large $N$.

Let us now see which measurement could be used to estimate the field gradient  with a totally polarized state.
The homogeneous field rotates all spins by the same angle, while the gradient rotates the spin at different positions by different angles.
Due to that, the homogeneous field rotates the collective spin, but does not change its absolute value. On the other hand, the field gradient decreases the absolute value of the spin since it has been prepared to be maximal, which has been used in Ref.~\cite{Behbood2013} for gradient magnetometry (see Fig.~\ref{fig:ionchain-evolution}).
Hence, we can measure the spin length to estimate the field gradient.

\subsubsection{Best separable state}
\label{sec:The best separable state}
We now turn our attention to the precision bound for all separable spin states.
It is useful to obtain this value so we have a direct comparison on what the best classically achievable precision is.
It turns out that for $j>\frac{1}{2},$ it is possible to achieve a precision higher than with the fully polarized state \eqref{eq:totally polarized state}.

Let us consider general separable states, which are not necessarily PI.
We do not know if the optimal separable state is sensitive or insensitive to the homogeneous field.
The corresponding precision bounds for the gradient estimation are given in Eqs.~\eqref{eq:bound-for-insensitive-and-thermal-state} and \eqref{eq:bound-for-sensitive-and-thermal-state}, respectively.
Since the probability density function \eqref{eq:pi-for-pdf} is PI, we have \be \int x_n P(x) dx = \mu \ee for all $n.$
As explained in Sec.~\ref{sec:Precision bound for states sensitive to homogeneous fields: Two-parameter dependence}, by moving the ensemble, the precision bounds do not change.
If we move the system to the origin of the coordinate system achieving $\mu=0,$
we can make our calculations simpler since the second term appearing in Eq.~\eqref{eq:bound-for-sensitive-and-thermal-state} is zero.
Thus, we only compute the first term in Eq.~\eqref{eq:bound-for-sensitive-and-thermal-state}.
Hence, the two bounds \eqref{eq:bound-for-insensitive-and-thermal-state} and \eqref{eq:bound-for-sensitive-and-thermal-state} are the same in this case and we arrive at
\be\label{eq:bound rewritten}
     \varinv{b_1}_{\text{sep}}|_{\max} = \sum_{n,m}\int x_n x_m P(\bs{x})\dif{\bs{x}} \qfif{\varrho^{(\text{s})}, j_z^{(n)}, j_z^{(m)}},
\ee
where we already assume that the bound can be saturated (see Appendix~\ref{app:compatibility-of-measurements}).

We now look for the separable state that maximizes the right-hand side of Eq.~\eqref{eq:bound rewritten}, which has to be a pure product state due to the convexity of the quantum Fisher information.
Hence, we look for the pure product state maximizing $\qfif{\varrho^{(\text{s})}, j_z^{(n)}, j_z^{(m)}}.$ Based on Eq.~\eqref{eq:fab-pure-states},
for product states we find that
\be\label{eq:sumFQnm_productstate}
  \qfif{\varrho^{(\text{s})}, j_z^{(n)}, j_z^{(m)}}=
  \begin{cases} 0&\text{if } n\ne m,\\
  4\varian{j_z^{(n)}} &\text{if } n=m.
  \end{cases}
\ee
For all $n$, a state that maximizes Eq.~\eqref{eq:sumFQnm_productstate} is
\be\label{eq:best separable state}
  \ket{\psi_{\text{sep}}} = \left(\tfrac{\ket{-j}+\ket{+j}}{\sqrt{2}}\rpar^{\otimes N},
\ee
for which the single-particle variances are maximal, i.e, $\varian{j_z^{(n)}}=j^2.$ While we carried out an optimization over general, non-necessarily PI separable states, the optimal state is PI. Plugging the state \eqref{eq:best separable state} into the bound given in Eq.~\eqref{eq:bound rewritten}
leads to  the precision bound for separable states as
\be\label{eq:bound-separable}
\begin{split}
    \varinv{b_1}_{\text{sep}} |_{\max} = 4\sigma^2 N j^2,
\end{split}
\ee
where we have used the definition of the variance of the particle positions \eqref{eq:variance} for a permutational invariant state.

Note that the bound for the best separable state given in Eq.~\eqref{eq:bound-separable} is above the bound obtained for the singlet state \eqref{eq:sing_QFI}, whereas the bound for the totally polarized state in Eq.~\eqref{eq:precision bound for the totally polarized state} is below.
Nevertheless, when the singlet state is used, the homogeneous magnetic field has no effect on the state.
In contrast, the state \eqref{eq:best separable state} is sensitive to the homogeneous field.

\subsubsection{Unpolarized Dicke states $\ket{{\rm {D}}_N}$ and $\ket{{\rm D}_{N}}_x$}

Next, we compute precision bounds for entangled states. In this section, we consider unpolarized Dicke states, which play an important role in quantum optics and quantum information science.
The Dicke state $\ket{\dicke{N}}_l$ [Eq.~\eqref{eq:dicke-def}], with a maximal $\langle J_x^2+J_y^2+J_z^2 \rangle$ and $\langle J_l\rangle=0$ for any $l\in x,y,z$ is particularly interesting due to its entanglement properties and its metrological usefulness \cite{Dicke1954,Toth2007}.
This state has been created in photonic experiments \cite{Kiesel2007,Wieczorek2009,Chiuri2012} and in cold atoms \cite{Luecke2011,Hamley2012}, while a Dicke state with $\langle J_z\rangle>0$ has been created with cold trapped ions \cite{Haeffner2005}.

The Dicke state $\ket{\dicke{N}}$ is an eigenstate of $J_z$ so it is insensitive to a homogeneous magnetic field pointing into the $z$ direction. Thus, the precision bound can be saturated by some measurement.
The Dicke state $\ket{\dicke{N}}_x$ is sensitive to the homogeneous field.
Moreover, it is very useful for estimating the homogeneous field as it has been shown in Ref.~\cite{Luecke2011}. Here, we consider large particle numbers, to make the results simpler.

Let us now see the most important properties of Dicke states.
For the expectation values of the single-particle angular momentum components
\bea\label{eq:meanjzn_Dicke}
  \meanO{j_l^{(n)}}&=&0
  \eea
  hold for $l=x,y,z$ for all $n.$
The second moments of the collective angular momentum components are given as
\bea
  \meanO{J_x^2}=\meanO{J_y^2}=\frac{N}{4} \left(\frac{N}{2}+1\right), \quad
  \meanO{J_z^2}=0.
\eea
Let us now see two-body correlations. Since the Dicke state is PI, we have
\bea
 \meanO{j_l^{(n)}j_l^{(m)}}=\meanO{j_l^{(1)}j_l^{(2)}},\quad
  \meanO{(j_l^{(n)})^2}=\meanO{(j_l^{(1)})^2}
\eea
for all $m\ne n$ and $l=x,y,z.$ Hence, the collective second moments are connected to the single particle and two-particle operator expectation values as
\be
    \meanO{J_l^2}=N\meanO{(j_l^{(1)})^2}+N(N-1)\meanO{j_l^{(1)}j_l^{(2)}}
\ee
for $l\,{=}\,x,y,z.$
Considering the symmetry under rotations around $z$ axis, we also have
$\meanO{(j_{x}^{(1)})^2}\,{=}\,\meanO{(j_{y}^{(1)})^2}$,
$\meanO{j_{x}^{(1)}j_{x}^{(2)}}\,{=}\,\meanO{j_{y}^{(1)}j_{y}^{(2)}}.$
Based on these and using Eq.~(\ref{eq:jx2jy2jz2}) for $j=1/2,$ we arrive at \cite{Urizar-Lanz2013}
\be
  \meanO{(j_l^{(n)})^2}= \frac{1}{4}\label{eq:meanjzn2_Dicke}
\ee
for $l=x,y,z$.

After discussing the main properties of the Dicke states, we can now obtain a precision bound for gradient metrology with such states.

\begin{observ}
For large $N$, the precision bound for the Dicke state $\ket{\dicke{N}}$ is
\be
  \label{eq:exact precision bound for dicke ih state}
  \varinv{b_1}_{\dicke{}}|_{\max} = (\sigma^2-\eta) N.
\ee
For the  Dicke state $\ket{\dicke{N}}_x,$ the  precision is bounded as
\be\label{eq:bound for dicke sh state}
  \varinv{b_1}_{\dicke{},x} |_{\max} = (\sigma^2 -\eta) N + \eta \frac{N(N+2)}{2},
\ee
which allows in principle a Heisenberg limited behavior due to the second term on
the right-hand side.
\label{obs:bound-ih-sh-dicke}
\end{observ}

\begin{proo}
Let us prove first Eq.~\eqref{eq:exact precision bound for dicke ih state}.
Since $\ket{\dicke{N}}$ is a pure state,
the QFIs appearing in Eq.~\eqref{eq:precision bound for single-cloud
systems IH} are simply four times the corresponding variances of $j_z^{(n)}$.
Based on the relations \eqref{eq:meanjzn_Dicke} and \eqref{eq:meanjzn2_Dicke} giving the first and second moments of $j_l^{(n)},$ respectively,
we obtain
\be
\qfif{\ket{\dicke{N}},j_z^{(n)}} = 4 (\Delta j_z^{(n)^2})=1.\label{eq:FQ1}
\ee
From Eq.~\eqref{eq:FQ1} and the bound for states insensitive to the homogeneous field \eqref{eq:precision bound for single-cloud
systems IH},
the precision bound for the Dicke state $\ket{\dicke{N}}$ follows.

We prove now the bound for the $\ket{\dicke{N}}_{x}$ states given in Eq.~\eqref{eq:bound for dicke sh state}.
The second moments $\meanO{(j_z^{(n)})^2}$ for $\ket{\dicke{N}}_{x}$ can be obtained from the second moments computed above for $\ket{\dicke{N}}$ by relabeling the coordinate axes.
Since $\ket{\dicke{N}}_x$ is a pure state,
the QFI again equals four times the corresponding variance.
Hence, we obtain
\begin{equation}
 \begin{split}
  \qfif{\ket{\dicke{N}}_x,j_z^{(n)}} & =  1, \\
  \qfif{\ket{\dicke{N}}_x,J_z}& = N(N+2)/2,
 \end{split}
\end{equation}
 and using the bound for states sensitive to homogeneous fields given in Eq.~\eqref{eq:precision bound for single-cloud systems SH} we have all we need to prove Observation~\ref{obs:bound-ih-sh-dicke}.
\end{proo}

\subsubsection{GHZ state}

The GHZ states are defined for qubits in Eq.~\eqref{eq:ghz-def}.
Such states are very sensitive to the homogeneous field.
GHZ states are highly entangled and play an important role in quantum information theory \cite{Greenberger1989}.
They have been created experimentally in photonic systems \cite{Pan2000, Yao2012, Lu2007} and trapped ions \cite{Sackett2000, Monz2011}.

Let us see first the relevant expectation values for GHZ states. Direct calculation shows that
\bea
\ex{j_z^{(n)}}=0,\quad
\ex{J_z}=0.
\eea
Moreover, for the second moments
\bea
\meanO{(j_z^{(n)})^2}{=}\frac{1}{4},\quad
\meanO{J_z^2} = \frac{N^2}{4}
\eea
hold.

Let us now calculate the precision bound. We recall that for pure states the QFI is given as Eq.~\eqref{eq:fab-pure-states}.
Using the bound for states sensitive to homogeneous fields given in Eq.~(\ref{eq:precision bound for single-cloud systems SH}),  we obtain
\be\label{eq:precision bound for ghz}
  \varinv{b_1}_{\ghz}|_{\max} = (\sigma^2-\eta) N + \eta N^2.
\ee
From \eqref{eq:precision bound for ghz} follows that we can reach the Heisenberg limit with such states, but only in
cases where $\eta$ is positive, i.e., when the particles are spatially correlated.

\vspace{-0.8em}
\subsubsection{Summary of results}
\vspace{-0.5em}

\begin{figure}[!b]
\begin{center}
\includegraphics[width=210pt]{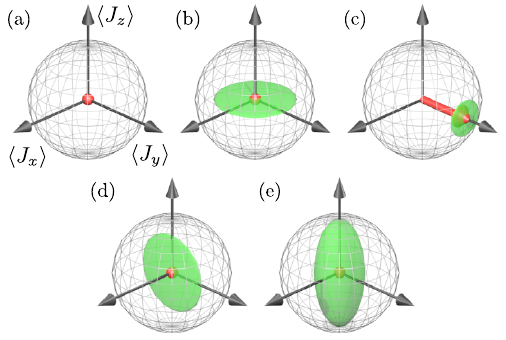}
\caption{Angular momentum components and their variances for
various spin states for few particles are shown.  (a) Singlet state, (b) Dicke state, (c) state totally polarized in the $y$ direction, (d) Dicke state in the $x$ direction, and (e) the GHZ state.
(Red vector) Angular momentum components $(\ex{J_x},\ex{J_y},\ex{J_z})$. (Green ellipse) Variances of the angular momentum components.
The radius of the sphere is the maximal angular momentum $r=Nj$.
}
\label{fig:different states}
\end{center}
\end{figure}

Finally, we summarize the precision bounds obtained for various quantum states in Table~\ref{tab:compare all the states}.
In Fig.~{\ref{fig:different states}}, we show the mean values and variances
of the collective angular momentum components for these states.
Note that for these PI states the optimal estimators for the homogeneous field and the gradient field are compatible (see  Appendix~\ref{app:compatibility-of-measurements}).
It means that the two parameters can be estimated at once even for the states sensitive to the homogeneous fields.

\begin{table}
  \caption{Precision bounds for  differential magnetometry for various quantum states defined in the main text.}
  \label{tab:compare all the states}
  \renewcommand\arraystretch{1.4}
  \begin{tabular}{
    m{0.25\linewidth}
    >{\centering\arraybackslash}m{0.4\linewidth} }
  \hline
  \hline
  States & $\varinv{b_1}|_{\max}$\\
  \hline
  $\varrho_{\text{singlet}}$ & $(\sigma^2-\eta) N 4j(j+1)/3$\\
  $\ket{j}_y^{\otimes N}$ & $ 2\sigma^2 N j$ \\
  $\ket{\psi_{\text{sep}}}$ & $ 4\sigma^2 N j^2$ \\
  $\ket{\dicke{N}}$ & $(\sigma^2-\eta) N$\\
  $\ket{\dicke{N}}_{x}$ & $
  (\sigma^2 -\eta)N + \eta N(N{+}2)/2$ \\
  $\ket{\ghz}$ & $ (\sigma^2-\eta) N + \eta N^2$ \\
  \hline
\end{tabular}
\end{table}

\vspace{-0.5em}
\section{Gradient Magnetometry with a Bose-Einstein Condensate}
\label{sec:bec}
\vspace{-0.5em}

In this section we study the case when our external state is a Bose-Einstein condensate instead of an incoherent mixture of pointlike particles.
We can write the spatial state of a BEC [Eq.~\eqref{eq:BEC}] as
\begin{equation}
    \varrho^{(\text{x})}_{\rm BEC} = \ketbra{0}{0},
\end{equation}
where we define the state $\ket{0}$ as the pure product state representing the BEC.

Since all particles are in the same spatial state, several important quantities describing the ensemble
can easily be computed. For such a quantum state,  for the average particle position defined in Eq.~\eqref{eq:mean} we obtain
\be
\mu =\braOket{0}{x^{(n)}}{0}\label{eq:muBEC}
\ee
for all $n.$
For the variance given in Eq.~\eqref{eq:variance},
\be
\sigma^2=\braOket{0}{(x^{(n)})^2}{0} - \mu^2\label{eq:sigma2BEC}
\ee
holds for all $n.$ Finally, there is no correlation between particle positions, i.e.,  $\langle x^{(n)}x^{(m)}\rangle=\langle x^{(n)}\rangle\langle x^{(m)}\rangle$ if $n\ne m.$ Hence,
the covariance, \eqref{eq:covariance}, is zero
\be
\eta=0.
\ee
Finally, as explained in Sec.~\ref{sec:Precision bound for states sensitive to homogeneous fields: Two-parameter dependence}, the precision bounds do not change if we translate the system. We move the atomic ensemble to the origin of the coordinate system such that\be
\mu=0.\label{eq:Muis0}
\ee
This will make our calculations much simpler.

Based on Eq.~\eqref{eq:one-parameter-precision}, for states insensitive to the homogeneous field we obtain
\be\label{eq:BEC_ins}
(\Delta b_1)^2\vert_{\max} = \mathcal{F}_{11}.
\ee
Based on Eq.~\eqref{eq:precision bound for b1 in terms of QFI matrix elements}, for states sensitive to the homogeneous field we obtain
\be\label{eq:BEC_sens}
(\Delta b_1)^2\le\mathcal{F}_{11}.
\ee
Here we used that
\be
\mathcal{F}_{10}=\mathcal{F}_{10}=4\mu (\Delta J_z)^2
\ee
is zero due to Eq.~\eqref{eq:Muis0}.
The bounds needed in Eqs.~\eqref{eq:BEC_ins} and \eqref{eq:BEC_sens} are equal to each other and can be obtained as follows.
We will compute a bound on $\mathcal{F}_{11}$ on pure states. Straightforward algebra leads to
\be
\mathcal{F}_{11}=4(\Delta H_1)^2=4\sigma^2 \tr\left[\textstyle{\sum_{n} (j_z^{(n)})^2 \varrho^{(\text{s})}}\right].
\ee
One can see that the optimal spin state for gradient estimation is the state totally polarized in the $z$ direction
\be\label{eq:optBEC}
\ket{\Psi}_{\rm opt,BEC}=\ket{j}^{\otimes N},
\ee
which is separable.
Hence, the precision is bounded for spin-$j$ particles as
\begin{equation}\label{eq:varinv_BEC}
    \varinv{b_1} |_{\max} = 4\sigma^2 N j^2.
\end{equation}
This is quite surprising, since under the dynamics coupling to the $z$ component of the spin and hence it rotates around the $z$ axis. One would naively expect that the optimal state is the state totally polarized in the $y$ direction \eqref{eq:totally polarized state} studied in Sec.~\ref{sec:Totally polarized state} for the case of cold atomic ensembles.
Due to the convexity of the quantum Fisher information, the bounds are also valid for the case of a mixed spin state.

Based on Eq.~\eqref{eq:varinv_BEC}, we see that the Heisenberg scaling cannot be reached in this case. Interestingly, this is true for any spatial wave function. For instance, if a single BEC is in a double-well potential, it still cannot have a scaling better than the shot-noise scaling in gradient estimation. In contrast, in Sec.~\ref{sec:twin cloud systems} we have seen that a Heisenberg scaling is possible in a double well, if two independent BECs are in the two wells.

\vspace{-0.5em}
\vspace{-0.5em}
\section{Conclusions}
\vspace{-0.5em}

In this work, we investigated the precision limits of measuring the  gradient of a magnetic field with atomic ensembles arranged in different geometries and initialized in different states.
We were particularly interested as to how the best achievable precision scales with the number of particles.
For spin chains and the two-ensemble case, the precision of the  estimation of the gradient can reach the Heisenberg limit.
For a single ensemble with localized particles,
the shot-noise limit can be surpassed and even the Heisenberg limit can be achieved
if there is a strong correlation between the particle positions.
We also studied the case of a single Bose-Einstein condensate, and found that the shot-noise limit can not be surpassed  in
this case.
However, even if the Heisenberg limit is not reached,
single-ensemble methods can have a huge practical advantage compared to methods based on
two or more atomic ensembles since using a single ensemble makes the experiment simpler and can also result in a better spatial resolution.
Independently from our work, Ref.~\cite{Altenburg2017} studied gradient metrology for different configurations of $N$ particles distributed on a line. 
\vspace{-0.5em}

\begin{acknowledgments}
We thank J.~Calsamiglia, G.~Colangelo, R.~Demkowicz- Dobrza\'nski, I.~L.~Egusquiza,
O.~G\"uhne, S.~Altenburg, S.~W\"olk, M.~Oszmaniec, C.~Klempt, M.~W.~Mitchell, M.~Modugno,
L.~Santos, R.~J.~Sewell, and A.~Smerzi for stimulating discussions.
We acknowledge the financial support of the EU (ERC
Starting Grant No. 258647/GEDENTQOPT,
CHIST-ERA QUASAR,
COST Action CA15220, QuantERA CEBBEC), the Spanish Ministry of Economy, Industry and Competitiveness,
and the European Regional Development Fund FEDER
through Grant No. FIS2015-67161-P (MINECO/FEDER),
the Basque Government (Project No. IT986-16),
the UPV/EHU program UFI 11/55,
and the National Research, Development and Innovation Office NKFIH (Contracts No. K124351, No. K124152, and No.K124176).
 I. U.-L. acknowledges the support of
a Ph.D. grant of the Basque Government.
Z. Z. was supported by the J\'anos Bolyai Scholarship of the Hungarian Academy of Sciences.
\end{acknowledgments}

\appendix

\vspace{-0.5em}
\section{The effects of the movement\\ of the atoms on the precision}
\vspace{-0.5em}
\label{app:experimental-suport-for-omitting-bireANDthermal}

In this paper, we compute the precision bounds neglecting the displacement of the particles generated by the gradient field and the thermal dynamics of the particles.
We now first analyze the displacement induced by the gradient of the magnetic field, and next we analyze which are the blurring effects caused by the thermal dynamics.

First of all, let us assume that we have for the internal subspace a completely mixed $N$-particle state $\varrho^{(\text{s})}$ placed in a single point in space (see Fig.~\ref{fig:cloud-in-sg}).
From the famous experiment of Gerlach and Stern \cite{Gerlach1922}, we know that the final state is split in two.
Moreover, the more distance between the two final subensembles, the larger the gradient of the field.
Hence, surprisingly, taking into account the movement of the particles induced by the gradient reduces the error in the estimation, so neglecting it, our bounds on the precision are still valid.

Nevertheless,
the gradient induces a force which depends on the spin state of the atoms.
The force is constant, thus the position will change quadratically in time.
On the other hand, the spin state changes linearly.
Hence, for small enough evolution times the displacement of the particles can be neglected.

Moreover, in a typical experiment for sensing the gradient of the magnetic field, sensitivities of the order of \units{1}{pT/\textmu{}m} can be reached, for a gradient of the magnetic field of \units{100}{nT/\textmu{}m} \cite{Behbood2013, Behbood2014, Kubasik2009}.
Hence, the classical acceleration due to the gradient of the magnetic field is $a \approx g_{F} \mu_{\text{B}} B_1 / m$, where $m$ and $g_{F}$ are the mass and the gyromagnetic $g$-factor of a $^{87}$Rb atom, respectively,  $m\approx \units{87}{u}$ and $g_{F}\approx 0.5$.
This results in an acceleration of the order of $\units{3\times10^{-2}}{m/s$^2$}$.
After \units{0.5}{ms} of evolution \cite{Behbood2013}, the atom travels a distance of the order of \units{10}{nm}, which is irrelevant compared with the size of these systems.

Next, let us consider the thermalization of the state which introduces random displacements of the particles potentially blurring the signal.
A typical cigar-shape ensemble of $^{87}$Rb atoms used for gradientometry is a couple of millimeters long and temperatures around \units{20}{\textmu{}K}  \cite{Kubasik2009, Behbood2014}.
We use the formula that connects the mean-root-average velocity of the particles and the temperature, $\bar{v} = \sqrt{3k_{\text{B}}T/m}$.
Note that not all the particles move towards the same direction but randomly in any direction.
Hence, we compute the average of the modulus of the projection of the velocity parallel to the direction of the cloud as $\overline{|v_{\parallel}|} = \bar{v}/2$.
We conclude
that the atoms are displaced by around \units{19}{\textmu{}m} along the axis to the cloud, which again is
irrelevant for clouds of the size of millimeters \cite{Kubasik2009, Behbood2013}.

Moreover, the displacement due to the gradient and thermal dynamics can be clearly neglected
in the cases of the spin chain, the two  ensembles, and the BEC, which are discussed in Secs.~\ref{sec:twin cloud systems} and \ref{sec:bec}.
Hence, the precision bounds computed in this paper can be used as a tool to characterize different states.

Concerning the sensitivity of our magnetometer, we can say the following. Assuming $N=8.5\times10^{6}$ atoms, trap length $\sigma=3$~mm, and for the completely polarized state discussed in Sec.~\ref{sec:Totally polarized state}, we obtain $\Delta B_1 \approx \units{3}{pT/mm}$, which is similar to the state of the art of other cold gas magnetometers \cite{Behbood2013}. The precision can be considerably improved if we use entangled states and we have correlation between the particle positions. There are other setups that work at much lower length scales, however, it is difficult to compare them to our system since they would not work at mm length scales \cite{Behbood2013}.

\vspace{-0.7em}
\section{Spatial state of thermally distributed pointlike particles}
\label{app:thermal-state}
\vspace{-0.5em}

We discuss the spatial state represented by Eq.~\eqref{eq:thermal-state}. For that, let us introduce the position operator as
\be\label{eq:xhat-definition}
    \hat{\bs{x}} = \int \bs{x} \ketbra{\bs{x}}{\bs{x}} \dif \bs{x},
\ee
where $\bs{x}$ is a vector of the particle positions,
and $\ket{\bs{x}}$ denotes a spatial state in which the pointlike particles are at given positions with the usual normalization
\be
    \braket{\bs{x}}{\bs{y}}=\delta(\bs{x}-\bs{y}),
\ee
as expected.
Based on Eq.~\eqref{eq:xhat-definition}, we see that
\be
    \hat{\bs{x}}\ket{\bs{x}}=\bs{x}\ket{\bs{x}}.
\ee
Thus, $\ket{\bs{x}}$ is an eigenstate of the operator $\hat{\bs{x}}.$
In order to obtain a quantum state that represents $N$ pointlike particles placed in the locations determined by the $\bs{x}$ vector,
we have to normalize it as
\be
    \ket{\varphi_{\bs{x}}} = \frac{\ket{\bs{x}}}{\sqrt{\braket{\bs{x}}{\bs{x}}}}.
    \label{eq:point-like-particles-pure-state}
\ee
From Eq.~\eqref{eq:point-like-particles-pure-state} and using that there is a probability distribution function, $P(\bs{x})$, and defining $P(\bs{x})$ as the probability to find particles at a given position $\bs{x}$, we arrive at Eq.~\eqref{eq:thermal-state}.

\vspace{-0.8em}
\section{Calculation of the QFI matrix elements for pointlike particles}
\label{ap:long-calc}
\vspace{-0.5em}

In this appendix, we show how to compute the QFI $\qfif{\varrho,H_i,H_j }$ if the spatial part of the state is written as Eq.~\eqref{eq:thermal-state}.

Let us write first the density matrix in its eigenbasis as
\be
\begin{split}
  \varrho=&\int \frac{P(\bs{x})}{\braket{\bs{x}}{\bs{x}}} \ketbra{\bs{x}}{\bs{x}} \dif{\bs{x}} \otimes\sum_\lambda p_\lambda \ketbra{\lambda}{\lambda}\\
  =&\int\sum_\lambda \frac{P(\bs{x})p_\lambda}{\braket{\bs{x}}{\bs{x}}} \ketbra{\bs{x},\lambda}{\bs{x},\lambda}\dif{\bs{x}},
\end{split}
\ee
where $P(\bs{x})p_\lambda/\braket{\bs{x}}{\bs{x}}$ are the eigenvalues.
Based on Eq.~(\ref{eq:fab}), the QFI matrix elements are written as
\begin{multline}
  \qfif{\varrho,H_i,H_j }=  2\int \sum_{\lambda,\nu} \frac{1}{\braket{\bs{x}}{\bs{x}}} \frac{[P(\bs{x})p_\lambda-P(\bs{y})p_{\nu}]^2} {P(\bs{x})p_\lambda+P(\bs{y})p_{\nu}} \\
  \times(H_i)_{\bs{x},\lambda;\bs{y},\nu}(H_j)_{\bs{y},\nu;\bs{x},\lambda} \dif{\bs{x}}\dif{\bs{y}}.
  \label{eq:fij-original}
\end{multline}
Note that $\braket{\bs{x}}{\bs{x}}\equiv\braket{\bs{y}}{\bs{y}}$ and that the integral is over $2N$ variables, $\bs{x}$ and $\bs{y}$.

We now use the fact that the generators $H_0$ and $H_1$ are diagonal in the spatial basis [see Eqs.~\eqref{eq:generator0-diagonal-spatially} and \eqref{eq:generator1-diagonal-spatially}].
Hence, the matrix elements can be rewritten as
\be
    (H_i)_{\bs{x},\lambda;\bs{y},\nu} \equiv \delta(\bs{x}-\bs{y}) (\mathcal{H}_i)_{\lambda,\nu}
    \label{eq:hi-diagonal-in-position}
\ee
for $i\,{=}\,0,1$, where $\mathcal{H}_i$ is a shorthand for $\sum_{n=1}^N j_z^{(n)}$ and $\sum_{n=1}^N x_n j_z^{(n)}$, respectively.
Using $\braket{\bs{x}}{\bs{y}}\,{=}\,\delta(\bs{x}-\bs{y})$ and Eq.~\eqref{eq:hi-diagonal-in-position}, we write Eq.~\eqref{eq:fij-original} as
\be
\begin{split}
    \qfif{\varrho,H_i,H_j }= & \,
    2 \int \sum_{\lambda,\nu} P(\bs{x}) \frac{(p_\lambda-p_{\nu})^2} {p_\lambda+p_{\nu}}\\
    &\times(\mathcal{H}_i)_{\lambda,\nu}(\mathcal{H}_j)_{\nu,\lambda} \dif{\bs{x}},
    \label{eq:}
\end{split}
\ee
which using the definition \eqref{eq:fab} for $\qfif{\varrho^{(\text{s})}, j_z^{(n)}, j_z^{(m)}}$ simplifies to Eqs.~\eqref{eq:bound-for-insensitive-and-thermal-state}, \eqref{eq:f11}, \eqref{eq:f00}, and \eqref{eq:f01} depending on the case.

\vspace{-0.5em}
\section{Optimal measurements for singlet states}
\label{app: Optimal measurements for singlet states}
\vspace{-0.5em}

In this appendix, we prove that the precision limits for gradient metrology
can be saturated for singlet states if we measure $J_x^2.$

\begin{observ} Let the initial spin state of an atomic ensemble be an arbitrary PI singlet state $\varrho^{(\text{s})}_{\text{singlet}}$.
Consider the experimental setup when $b_1$ is obtained by measuring $J_x^2$.
The precision of estimating $b_1$, which is given by the error propagation formula, is optimal in the short-time limit, i.e.,
\be
\label{eq: Jx2_acc_short}
\lim_{t \to 0} \frac{|\partial_{b_1}\ex{J_x^2(t)}|^2}{\ex{J_x^4(t)}-\ex{J_x^2(t)}^2} = \qfif{\varrho^{(\text{s})}, H_1},
\ee
where $J_x^k(t)=U^\dagger(t) J_x^k U(t)$, the time-evolution unitary operator is of the form $U(t)=e^{-ib_1H_1}$, and $H_1$ is defined in Eq.~\eqref{eq:gradient shift generator}.
\end{observ}
\begin{proo}
Since for any pure singlet
\be\label{eq:singletJx2}
J_x^k\ket{0,0,D}=0,
\ee
holds [Eq.~\eqref{eq:definition of a general singlet}],
we have that
$\ex{J_x^2(0)}=\ex{J_x^4(0)}=0$.
For the numerator, we have
\begin{equation}
 \begin{split}
  \lim_{t \to 0} | \partial_{b_1}\ex{J_x^2(t)}|^2 = \lim_{t \to 0}
  \trcua{\partial_{b_1}[e^{i b_1 H_1} J_x^2 e^{-i b_1 H_1}]
  \varrho^{(\text{s})}}\Big|^2 \\
  =\Big| \trcua{i  H_1  J_x^2 \varrho^{(\text{s})}} -
  \trcua{iH_1 \varrho^{(\text{s})} J_x^2}\Big|^2=
  0.
 \end{split}
\end{equation}
We see that both the numerator and denominator of the right-hand side
of Eq.~\eqref{eq: Jx2_acc_short} go to zero as $t \to 0$, thus the l'Hospital
rule can be used applying the derivative $\partial_{b_1}$ in both
the denominator and the numerator, which yields
\be
\lim_{t \to 0}\varinv{b_1}=
\lim_{t \rightarrow 0} \frac{2\ex{\partial^2_{b_1}
J^2_x(t)} \ex{\partial_{b_1}J^2_x
(t)}}{\ex{\partial_{b_1} J_x^4(t)}-2\ex{J_x^2(t)}
\ex{\partial_{b_1} J_x^2(t)}}.
\ee
However, here the numerator and the denominator are again zero at $t=0$, so
we employ the l'Hospital rule once again and obtain
\begin{multline}
\label{eq:limit of the precision for the singlet}
\lim_{t \to 0}\varinv{b_1}= \\
\begin{split}
=&\lim_{t \rightarrow 0}  \frac{2\ex{\partial^2_{b_1}J_x^2(t)}^2 +
2\ex{\partial^3_{b_1}J_x^2(t)} \ex{\partial_{b_1}J_x^2(t)}}
{\ex{\partial_{b_1}^2 J_x^4(t)}-2\ex{\partial_{b_1} J_x^2(t)}^2-
\ex{J_x^2(t)}\ex{\partial_{b_1}^2 J_x^2(t)}}\\
=&\lim_{t \rightarrow 0}  \frac{2\ex{\partial^2_{b_1}J_x^2(t)}^2}
{\ex{\partial_{b_1}^2 J_x^4(t)}} =\frac{2\ex{\lcua H_1, [H_1, J_x^2]\rcua}^2} {\ex{\lcua H_1, [H_1,J_x^4]\rcua}}\\
=&\frac{4\ex{ H_1 J_x^2 H_1}^2} {\ex{ H_1 J_x^4 H_1 }},
\end{split}
\end{multline}
where we simplified the expectation values that are 0 and we used the Heisenberg equation of motion twice for the second derivatives and simplified the result, using Eq.~\eqref{eq:singletJx2} and the definition of the commutator, to rewrite the equation.

Next, we will compute the numerator and the denominator in Eq.~\eqref{eq:limit of the precision for the singlet}.
First of all using the angular momentum commutation relation $[j_z^{(n)}, j_x^{(m)}] = i\delta_{n,m}j_y^{(m)}$, we compute $[H_1, J_x]$ obtaining
\be
\label{eq:commutator of H1 and Jx}
[H_1, J_x] = \sum_{n = 1}^N x^{(n)}[j_z^{(n)}, J_x] = i\sum_{n = 1}^N x^{(n)}j_y^{(n)} =: i H_{1,y}.
\ee
From the formula $[A, B^k] = \sum_{\alpha = 1}^{k}B^{\alpha -1} [A,B] B^{k -\alpha}$, and using Eq.~\eqref{eq:commutator of H1 and Jx}, we arrive at
\be\label{eq:commutator H1 and Jxk}
[H_1,J_x^k] = i\sum_{\alpha = 1}^{k} J_x^{\alpha-1} H_{1,y} J_x^{k-\alpha},
\ee
and similarly,
\be\label{eq:commutator H1y and Jxk}
[H_{1,y},J_x^k] = -i\sum_{\alpha = 1}^{k} J_x^{\alpha-1} H_{1} J_x^{k-\alpha}.
\ee
Now, using the commutator relations \eqref{eq:commutator H1 and Jxk} and \eqref{eq:commutator H1y and Jxk}, and  Eq.~\eqref{eq:singletJx2}, we are able to substitute $H_1 J_x^k$ by $[H_1,J_x^k]$ for which only remains the first term in the summation, $\alpha = 1$, and repeating the procedure for $H_{1,y}J_x^{k-1}$, we obtain
\be\label{eq:Jx2H1}
  \ex{H_1 J_x^k H_1} = i\ex{H_{1,y}J_x^{k-1} H_1}
   = \ex{H_{1}J_x^{k-2} H_1}.
\ee
Hence, we have that $\ex{H_1 J_x^k H_1} = \ex{H_1^2}$ for any even $k$.
Finally, from Eq.~\eqref{eq:limit of the precision for the singlet}, we arrive at $\lim_{t \to 0}\varinv{b_1}=4\ex{H_1^2}$ which for the case of the singlets is equal to $4\varian{H_1}$ since $\ex{H_1} = 0.$ Hence, the proof follows.
\end{proo}

\section{Proof that the precision bounds can be saturated}
\label{app:compatibility-of-measurements}

When working with a state that is sensitive to the homogeneous field,
in order to optimally estimate the gradient, one must measure simultaneously the gradient and the homogeneous field.
In other words, the optimal measurement for the homogeneous field and for the
gradient parameter should commute with each other.
In this section, we will show that in all cases we considered the two measurements
commute with each other. As a consequence, our bounds on the
precision obtained based on the formalism given in Sec.~\ref{sec:Precision bound for states sensitive to homogeneous fields: Two-parameter dependence} can be saturated.

In order to proceed, it is necessary to define
the symmetric logarithmic
derivative (SLD) $L(\varrho,A)$ which has the property that
\begin{equation}
    \frac{L(\varrho,A) \rho + \rho L(\varrho,A)}{2} = i[\varrho,A].
\end{equation}
and for a density matrix with an eigendecomposition of the form \eqref{eq:rhoeigendecomp} is given as
\be
    L(\varrho,A)= 2i\sum_{\lambda \neq \nu}\frac{p_\lambda-p_{\nu}}{p_\lambda+p_\nu} \braOket{\lambda}{A}{\nu}\ketbra{\lambda}{\nu}.
\ee

Then, quantum metrology tells us that the condition for being able to construct compatible
measurements to estimate $b_0$ and $b_1$ is \cite{Paris2009}
\be
\label{eq:condition for compatible measurements}
[L(\varrho,H_0),L(\varrho,H_1)]=0.
\ee
The two SLDs can be obtained as
\begin{subequations}
  \bea\label{eq:l0}
 L(\varrho,H_0) & =& \mtxid^{(\text{x})} \otimes L(\varrho^{(s)},J_z),\\
    \label{eq:l1}
    L(\varrho,H_1) & =& \sum_{n=1}^N  \int \dif{\bs{x}} \,x_n \ketbra{\bs{x}}{\bs{x}} \otimes L(\varrho^{(s)},j_z^{(n)}),
  \eea
  \end{subequations}
  after reordering the subspaces. For all cases when the
  internal state is permutationally invariant,
we arrive at the following expressions for the SLDs:
 \begin{subequations}
   \bea
    \label{eq:l0-pi}
    L(\varrho,H_0) & = \mtxid^{(\text{x})} \otimes L(\varrho^{(s)},J_z),\\\label{eq:l1-pi}
    L(\varrho,H_1) & = \hat{\mu}^{(\text{x})} \otimes L(\varrho^{(s)},J_z),
  \eea
  \end{subequations}
  where the SLD for the spin state is given as
   \be
  L(\varrho^{(s)},J_z) = 2i \sum_{\lambda,\nu} \frac{p_\lambda - p_\nu}{p_\lambda + p_\nu}
    \braOket{\lambda}{J_z}{\nu}
    \ketbra{\lambda}{\nu}
    \ee
  and the average position operator is defined as
  \be
  \hat{\mu}^{(\text{x})}=\frac{1}{N}\sum_n \int \dif{\bs{x}}\, x_n \ketbra{\bs{x}}{\bs{x}}.
  \ee
 One can see by inspection that the operators given in Eqs.~\eqref{eq:l0-pi} and
\eqref{eq:l1-pi} commute with each other.
Hence, all our bounds for PI states in Sec.~\ref{sec:single cloud systems} can be saturated, and all PI states discussed in other sections as well.

Finally, we have to discuss the states appearing in Table~\ref{tab:result-states-two-ensembles}.
They are product states of two PI states of $N/2$ particles each.
Thus, in terms of ``L'' and ``R,'' we have the following expression for $L(\varrho, H_1)$:
\begin{equation}
  \begin{split}
    L(\varrho, H_1) = &\, \hat{\mu}^{(\text{L})}\otimes \mtxid^{(\text{R},\text{x})}\otimes L(\ket{\psi}^{(\text{L})}, J_z^{(\text{L})}) \otimes (\ketbra{\psi}{\psi})^{(\text{R})} \\
    &+  \mtxid^{(\text{L},\text{x})}\otimes\hat{\mu}^{(\text{R})}\otimes
    (\ketbra{\psi}{\psi})^{(\text{L})} \otimes L(\ket{\psi}^{(\text{R})}, J_z^{(\text{R})}),
  \end{split}
\end{equation}
where $\hat{\mu}^{(\text{L})}$ is the average position operator for the ``L'' ensemble, similarly for $\hat{\mu}^{(\text{R})}$, and $J_z^{(\text{L})}$ is the $z$ projection of the total angular momentum of the ``L'' subsystem, as $J_z^{(\text{R})}$ is for ``R.''
Clearly, the operator $L(\varrho, H_1)$ commutes with $L(\varrho, H_0),$ which is given in  Eq.~\eqref{eq:l0-pi}.

With this, we conclude this appendix which let us demonstrate that all bounds in this paper can be saturated.

\bibliography{bibliography}

\end{document}